\documentclass{aa}  

\usepackage{graphicx}
\usepackage{txfonts}
\usepackage{lipsum}
\usepackage{subcaption}         
\usepackage{lscape}             
\usepackage{placeins}           
                                
\begin{document}

\title{Multi-tracer exploration of molecular gas in main sequence galaxies at $z\sim 4.5$}

\author{M. Dessauges-Zavadsky\inst{1},
            M. B\'ethermin\inst{2},
            M. Ginolfi\inst{3},
            L. Vallini\inst{4},
            A. Faisst\inst{5},
            M.-Y. Xiao\inst{1},
            F. Pozzi\inst{4,6},
            P. Cassata\inst{7,8},
            Y. Fudamoto\inst{9},
            C. Gruppioni\inst{4},
            G.~C. Jones\inst{10,11},
            M. Kohandel\inst{12},
            G. Rodighiero\inst{7,8},
            M. Romano\inst{13,8},
            P. Theul\'e\inst{14},
            G. Zamorani\inst{4},
            C. Accard\inst{2},
            \and 
            C. Guillaume\inst{2}
        }

\institute{D\'epartement d'Astronomie, Universit\'e de Gen\`eve, Chemin Pegasi 51, 1290 Versoix, Switzerland\\
\email{miroslava.dessauges@unige.ch}
\and
Universit\'e de Strasbourg, CNRS, Observatoire astronomique de Strasbourg, UMR 7550, 67000 Strasbourg, France
\and
Dipartimento di Fisica e Astronomia, Universit\`a di Firenze, Via G. Sansone 1, I-50019 Sesto F.no (Firenze), Italy
\and
INAF, Osservatorio di Astrofisica e Scienza dello Spazio, Via P. Gobetti 93/3, I-40129, Bologna, Italy
\and
IPAC, California Institute of Technology, 1200 E. California Blvd. Pasadena, CA 91125, USA
\and
University of Bologna, Department of Physics and Astronomy ``Augusto Righi'', Via Gobetti 93/2, 40129 Bologna, Italy
\and
Dipartimento di Fisica e Astronomia, Universit\`a di Padova, Vicolo dell’Osservatorio, 3, 35122 Padova, Italy
\and
INAF -– Osservatorio Astronomico di Padova, Vicolo dell’Osservatorio 5, 35122 Padova, Italy
\and
Center for Frontier Science, Chiba University, 1-33 Yayoi-cho, Inage-ku, Chiba 263-8522, Japan 
\and
Kavli Institute for Cosmology, University of Cambridge, Madingley Road, Cambridge CB3 0HA, UK
\and
Cavendish Laboratory, University of Cambridge, 19 JJ Thomson Avenue, Cambridge CB3 0HE, UK
\and
Scuola Normale Superiore, Piazza dei Cavalieri 7, 56126 Pisa, Italy
\and
Max-Planck-Institut für Radioastronomie, Auf dem Hügel 69, 53121 Bonn, Germany
\and
Aix-Marseille Univ., CNRS, CNES, LAM, Marseille, France
}

\date{}

\abstract{Molecular gas masses in high-redshift galaxies are inferred from indirect tracers, whose respective reliability remains poorly constrained. In particular, the bright [C\,{\sc ii}] 158~$\mu$m line is now widely used as a molecular gas tracer at $z>4$, yet direct observational tests against CO remain scarce.}
{We search for CO(4--3), CO(5--4), and [C\,{\sc i}](1--0) emission lines in three of the most [C\,{\sc ii}]-luminous galaxies at $z\sim 4.5$ selected from the ALPINE survey to assess the detectability of these lines in high-redshift main-sequence (MS) galaxies and to test the reliability of [C\,{\sc ii}] emission as a molecular gas tracer through the cross-comparison of molecular gas masses inferred from six tracers.}
{We obtained NOEMA observations of the CO and [C\,{\sc i}] lines and compared molecular gas masses inferred from CO(4--3), CO(5--4), [C\,{\sc i}](1--0), [C\,{\sc ii}] emission, dust continuum, and [C\,{\sc ii}]-based dynamical masses, adopting standard empirical calibrations and conversion factors from the literature.}
{We detect the CO(4--3) and CO(5--4) lines at high significance in the near-solar metallicity galaxy DC873756, obtain a tentative CO(4--3) detection in the merging system DC818760, and detect no CO emission in the half-solar metallicity galaxy VC5110377875. The [C\,{\sc i}] line remains undetected in all three galaxies, reflecting its intrinsic faintness. In DC873756, the molecular gas masses inferred from the six considered tracers agree within their $1\sigma$ uncertainties despite the different systematics inherent to each tracer. The agreement suggests that, at least for some near-solar metallicity MS galaxies at $z\sim 4.5$, the CO spectral line energy distribution (SLED) and Milky Way CO-to-H$_2$ conversion factor adopted for MS galaxies at cosmic noon remain applicable, and that mid-$J$ CO transitions trace a substantial fraction of the molecular gas reservoir. In contrast, the CO non-detection in VC5110377875 is consistent with the reduced CO detectability expected at lower metallicities. Only in DC818760, we find an inconsistency between the [C\,{\sc ii}]-based molecular gas mass and masses derived from the other tracers, indicating a [C\,{\sc ii}] excess possibly reflecting enhanced emission from shocks and/or diffuse ionized gas in merger-driven ISM conditions.}
{Although limited to three galaxies, this work provides a rare observational cross-validation of different molecular gas tracers in MS galaxies at $z\sim 4.5$. Our results support the use of [C\,{\sc ii}] emission to deliver a first-order estimate of the molecular gas mass in high-redshift MS galaxies, while highlighting possible limitations in mergers. We also find that mid-$J$ CO lines can be detected in near-solar metallicity MS galaxies at high redshift and used to trace the bulk of the molecular gas reservoir.}


\titlerunning{Multi-tracer exploration of molecular gas at $z\sim 4.5$}
\authorrunning{M. Dessauges-Zavadsky et al.}

\maketitle
 
\nolinenumbers 
 
\section{Introduction}
\label{sect:introduction}

Galaxies contributing to $\sim 90$\% of the cosmic star-formation rate (SFR) density lie on the star-forming MS relation between stellar mass ($M_{\rm stars}$) and SFR, whose normalisation increases with redshift \citep{Whitaker+12,Speagle+14,Schreiber+15,Lee+17}. The MS is in place out to the epoch of reionisation \citep{Faisst+20,Leslie+20,Khusanova+21}, as confirmed by James Webb Space Telescope (JWST) observations \citep{Fujimoto+23,Rinaldi+25,Cole+25,Simmonds+25}. MS galaxies are thought to grow steadily over several Gyr through cold gas accretion from the cosmic web, which sustains star formation and triggers intense starburst episodes \citep{Dekel+09,Lilly+13}.

The NOEMA and ALMA millimetre/submillimetre interferometers have played a major role in establishing the molecular hydrogen (H$_2$) gas mass ($M_{\rm molgas}$) census, the fuel for star formation, of MS galaxies. Up to $z\sim 3$, the molecular gas fraction increases with redshift and closely follows the rise of the specific SFR \citep{Bethermin+15,Dessauges+15,Genzel+15,Scoville+16,Scoville+17,Tacconi+18, Tacconi+20}. Similarly, the cosmic $M_{\rm molgas}$ density traces the evolution of the cosmic SFR density from $z=0$ to its peak at $z\sim 2$ \citep{Riechers+19,Decarli+20,Magnelli+20}. Beyond $z>4$, however, measurements of $M_{\rm molgas}$ in MS galaxies remain scarce \citep{Liu+19,Dessauges+20,Scoville+23,Aravena+24}, despite this being a key epoch in galaxy evolution during which galaxies assembled their stellar mass $10-100$ times faster than present-day galaxies and transitioned from primordial systems to chemically and dynamically mature galaxies. This scarcity mainly arises from the lack of reliable and accessible tracers of cold H$_2$ gas ($\ll 100~\rm K$), motivating the investigation of various molecular gas tracers in the literature.

Carbon monoxide ($^{12}$CO) rotational transitions are theoretically among the most effective tracers of cold molecular gas, as CO forms in carbon- and oxygen-enriched molecular clouds and is the most abundant molecule after H$_2$. Low-$J$ ($J\leq 3$) CO transitions, characterized by low excitation energies ($\lesssim35$~K) and low critical densities ($\sim200~\rm cm^{-3}$), trace the bulk of the cold, diffuse H$_2$ reservoir and are widely used in studies of nearby galaxies \citep{Bolatto+13}. Mid-$J$ ($J = 4-6$) CO transitions arise from warmer and denser gas ($\gtrsim 10^4-10^5~\rm cm^{-3}$) and correlate well with infrared (IR) luminosity, and hence SFR, due to their higher excitation energies \citep{Daddi+15,Vallini+18,Valentino+20}. At $z>4$, however, low-$J$ CO lines become increasingly difficult to detect because of their reduced contrast against the elevated cosmic microwave background (CMB) temperature \citep{daCunha+13,Zhang+16}. Mid-$J$ CO lines may therefore be more readily detectable in MS galaxies at these early epochs, both because they are less affected by the CMB and because the ISM conditions in compact high-redshift galaxies with elevated SFRs reach higher densities and stronger UV radiation fields \citep{Isobe+23,Topping+25,Harikane+25}. These conditions may shift the CO SLED toward mid-$J$ transitions, as predicted by \citet{Vallini+18}. The reliability of $M_{\rm molgas}$ estimates derived from CO often remains uncertain because both the CO SLED and the CO-to-H$_2$ conversion factor are poorly constrained, as they depend on the ISM temperature, density, and metallicity \citep{Bolatto+13,CarilliWalter13}. These quantities may also vary significantly across and within galaxies, such that different CO transitions can arise from different spatial regions. To date, a number of $z>4$ dusty star-forming galaxies (DSFGs) have been detected in CO, some of which may lie at the massive end of the MS \citep{Birkin+21,Liao+24,Frias+25,Brinch+25}. However, CO measurements of more representative MS galaxies at $z>4$ with $M{\rm stars}\lesssim 10^{10}~M_{\odot}$ remain extremely scarce, with only four CO detections reported so far \citep{Dodorico+18,Pavesi+19,Zavala+22,Cescon+26}, leaving the CO properties of typical star-forming galaxies at early cosmic times largely unconstrained.

The [C\,{\sc ii}] 158~$\mu$m fine-structure emission line, the primary coolant of gas heated by star formation, is typically the brightest feature in the rest-frame far-infrared (FIR) spectra of galaxies. It has emerged as a promising tracer of molecular gas following the discovery of an empirical correlation between [C\,{\sc ii}] luminosity ($L_{\rm [CII]}$) and $M_{\rm molgas}$ in both high-redshift ($z>2$) \citep[e.g.,][]{Zanella+18,Zhao+24} and nearby galaxies \citep[e.g.,][]{Accurso+17,Madden+20,Ramambason+24}. However, the [C\,{\sc ii}] emission originates from multiple ISM phases, including the ionised gas, the cold diffuse neutral medium, and photodissociation regions (PDRs) associated with the outer layers of giant molecular clouds \citep{Stacey+91,Madden+93,Wolfire+03,Wolfire+10,Wolfire+22,Cormier+15, DiazSantos+17}. Several simulation studies modelling [C\,{\sc ii}] emission from diffuse neutral gas, PDRs, and molecular clouds in high-redshift galaxies nevertheless support the use of [C\,{\sc ii}] as a tracer of molecular gas \citep{Pavesi+19,Vizgan+22, Pallotini+22,Schimek+24,Casavecchia+25,Vallini+15,Gurman+24,Vallini+25,Bhagwat+26, Accard+26}. Thanks to ALMA, the [C\,{\sc ii}] line is now routinely detected in galaxies at $z>4$ \citep[e.g.][]{Capak+15,Bethermin+20,Bouwens+22}. In particular, the ALPINE survey of MS galaxies at $z\sim4.5-6$ \citep{LeFevre+20,Bethermin+20,Faisst+20} showed that [C\,{\sc ii}]-based molecular gas masses agree with estimates derived from the dust continuum and cold-gas dynamical masses \citep{Dessauges+20}, providing statistical support for [C\,{\sc ii}] as a tracer of $M_{\rm molgas}$ in the early Universe.

Since dust acts as a catalyst for H$_2$ formation \citep[even at the elevated dust temperatures of high-redshift galaxies;][]{Grieco+23}, the dust continuum has naturally emerged as an alternative tracer of molecular gas, with the advantage of being more readily detectable than CO and [C\,{\sc ii}] emission lines \citep[e.g.][]{Scoville+16,Scoville+23,Liu+19,Wang+22,Zavala+26}. Dust masses can be estimated either by fitting the thermal FIR spectral energy distribution \citep[SED;][]{Bethermin+15,Kaasinen+19} or from the cold dust continuum in the Rayleigh-Jeans regime \citep[$\lambda > 250~\mu \rm m$;][]{Scoville+14,Scoville+16}. However, the $M_{\rm molgas}$ estimates derived from dust remain uncertain because both the dust temperature, often poorly constrained by sparsely sampled FIR SEDs, and the dust-to-gas mass ratio (DGR) are not well known. This is particularly true for high-redshift galaxies, for which the few available measurements indicate little evolution in the DGR from $z=0$ to $z\sim5.3$ \citep[][and references therein]{PoppingPeroux22}. Moreover, dust continuum detections in typical star-forming galaxies become increasingly rare toward the epoch of reionisation, with only one detection reported so far at $z=8.31$ \citep{Tamura+19,Bakx+20}, limiting the use of dust as a molecular gas tracer at the highest redshifts.

The atomic carbon lines [C\,{\sc i}](1--0) at 492.161~GHz and [C\,{\sc i}](2--1) at 809.344~GHz have also been proposed as tracers of the bulk molecular gas reservoir in galaxies, supported by modern PDR models including non-equilibrium chemistry \citep{Papadopoulos+04,Papadopoulos+18,Bisbas+15,Glover+15}. The [C\,{\sc i}] lines are optically thin and therefore probe higher column densities of cold molecular gas than $^{12}$CO, while avoiding the excitation bias affecting the $J\geq 4$ CO transitions. As discussed by \citet{Valentino+18}, models predict that the [C\,{\sc i}](1--0) luminosity correlates well with $M_{\rm molgas}$ over a wide range of ambient radiation fields and gas densities ($10-10^4~\rm cm^{-3}$). [C\,{\sc i}] detections have been reported in MS galaxies at $z\sim 1$ \citep{Valentino+18,Valentino+20,Bourne+19}, as well in numerous submilleter galaxies \citep[SMGs; see the compilation in][]{Dunne+22} and bright {\it Herschel}-selected galaxies \citep[z-GAL;][]{Berta+23} at $z=1-6$, and in South Pole Telescope (SPT) submillimeter dusty star-forming galaxies (DSFGs) at $z=1.8-4.8$ \citep{Gururajan+23}. Combined with CO observations, these measurements have been used to calibrate the abundance of atomic carbon relative to H$_2$. However, no [C\,{\sc i}] line has yet been detected in MS galaxies beyond cosmic noon.

\begin{figure}[!]
\centering
\includegraphics[width=0.45\textwidth,clip]{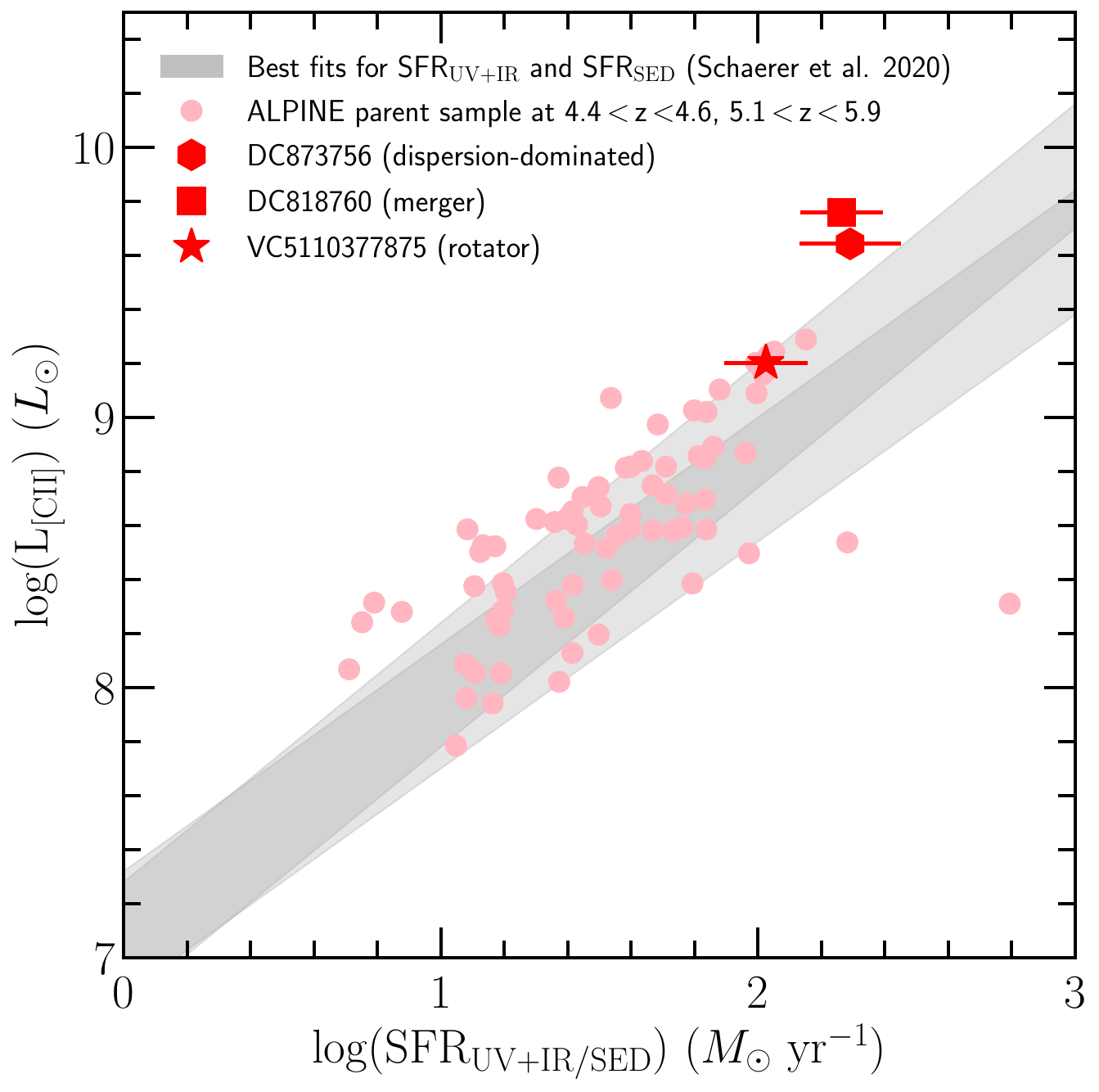}
\caption{[C\,{\sc ii}] luminosities as a function of SFRs of the ALPINE [C\,{\sc ii}]-detected galaxies (pink circles). The red symbols show the three galaxies selected for the CO(4--3), CO(5--4), and [C\,{\sc i}](1--0) NOEMA observations. They are among the most [C\,{\sc ii}]-luminous ALPINE galaxies and have different ISM morpho-kinematic properties ranging from dispersion-dominated (hexagon), merger (square) to rotator (star). The SFRs correspond to either the UV+IR-derived SFRs for galaxies detected in the ALMA dust continuum \citep{Bethermin+20} or the SED-derived SFRs \citep{Faisst+20}. The grey shaded bands show the $L_{\rm [CII]}$--SFR best fits obtained by \citet{Schaerer+20} for the ALPINE galaxies considering either $\rm SFR_{UV+IR}$ (steeper relation) or $\rm SFR_{SED}$ (shallower relation) with their respective $1\sigma$ dispersions.}
\label{fig:LCII-SFR}
\end{figure}
\begin{figure}[!]
\centering
\includegraphics[width=0.44\textwidth,clip]{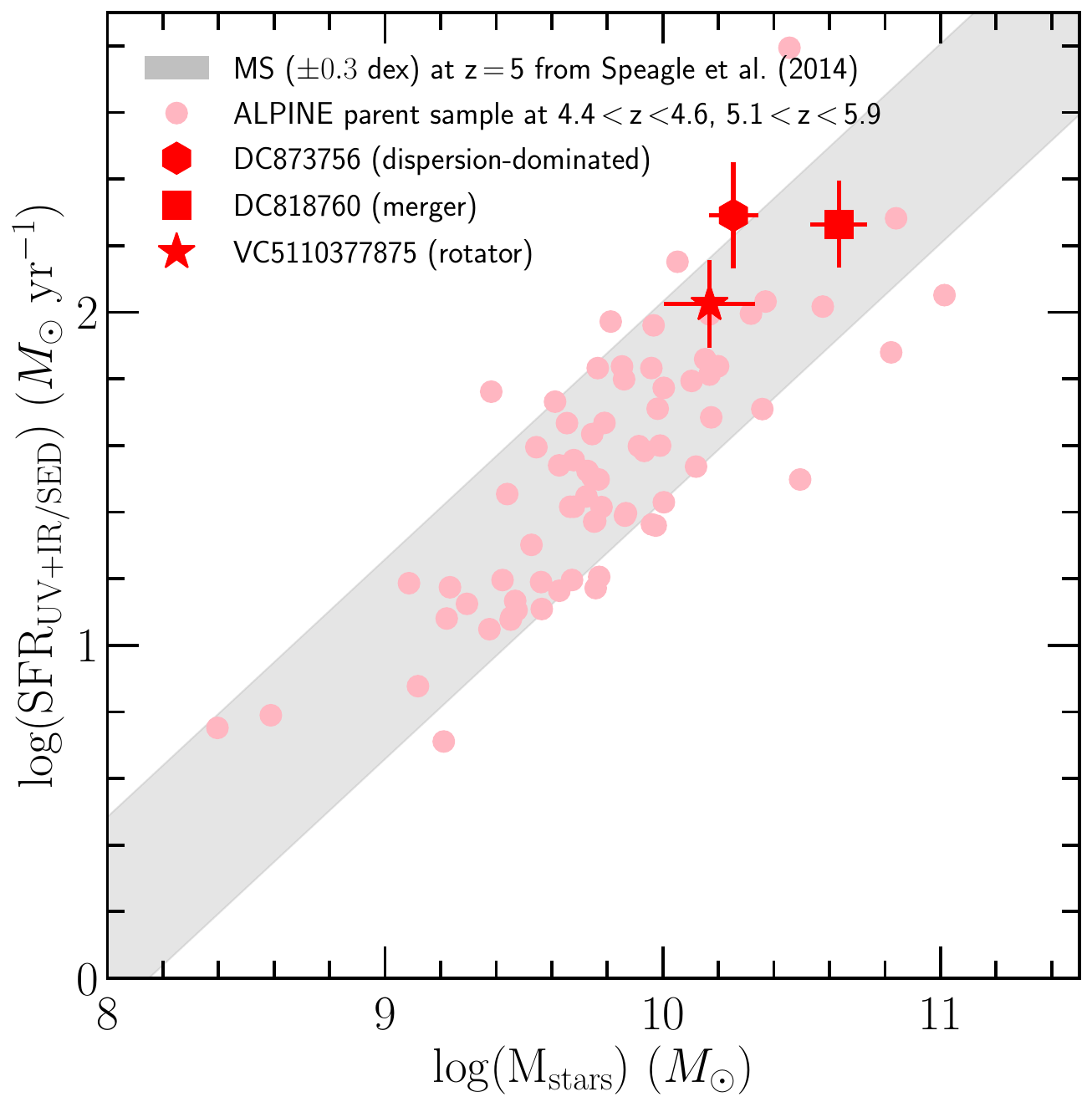}
\caption{Star formation rates as a function of stellar masses of the ALPINE [C\,{\sc ii}]-detected galaxies (pink circles). By selection the three studied [C\,{\sc ii}]-luminous galaxies are massive ($M_{\rm stars} >10^{10}~M_{\odot}$) and highly star-forming (${\rm SFR} > 100~M_{\odot}~\rm yr^{-1}$) found at the end of the ALPINE galaxy distribution (red hexagon, square, and star). The grey shaded band represents the main sequence of star-forming galaxies at $z=5$ as derived by \citet{Speagle+14} with a typical $\pm 0.3$~dex dispersion.}
\label{fig:MS}
\end{figure}

In this work, we present an exploratory analysis of the ISM molecular gas budget in a pilot sample of three bright [C\,{\sc ii}]-emitting MS galaxies at $z\sim4.5$, selected from the ALPINE survey and spanning different ISM morpho-kinematic properties, through simultaneous observations of multiple molecular gas tracers. Exploiting the 31~GHz bandwidth of NOEMA, we observed the CO(4--3), CO(5--4), and [C\,{\sc i}](1--0) emission lines in a single frequency setting for each galaxy. These observations complement existing ALMA data of [C\,{\sc ii}] and underlying rest-frame $\sim158~\mu$m continuum \citep{Bethermin+20,Dessauges+20,Devereaux+24,HerreraCamus+25}. Altogether, the NOEMA and ALMA datasets provide a rare opportunity to cross-compare all the molecular gas tracers described above in three galaxies representative of the bulk of the MS star-forming galaxy population at the end of the reionisation era. 

Section~\ref{sect:targets} describes the selection and physical properties of the three ALPINE galaxies. In Sect.~\ref{sect:observations} we present the NOEMA observations, data reduction, and imaging. The CO(4--3), CO(5--4), and [C\,{\sc i}](1--0) flux and luminosity measurements, together with the corresponding molecular gas mass estimates, are outlines in Sect.~\ref{sect:results}. In Sect.~\ref{sect:discussion} we compare six different tracers used to estimate the molecular gas masses and discuss the caveats affecting the CO and [C\,{\sc ii}] tracers. Sect.~\ref{sect:conclusions} summarizes the results.

Throughout the paper, we assume the $\Lambda$CDM cosmology with $\Omega_{\rm m} = 0.3$, $\Omega_{\Lambda} = 0.7$, and $H_0 = 70~\rm km~s^{-1}~Mpc^{-1}$. We adopt the \citet{Chabrier03} initial mass function.

\section{Target selection}
\label{sect:targets}

For the CO(4--5), CO(5--4), and [C\,{\sc i}](1--0) follow-up observations, we selected three of the most [C\,{\sc ii}]-luminous galaxies from the ALPINE survey, which targeted 118 UV-selected MS galaxies at $4.4<z<4.6$ and $5.1<z<5.9$, as illustrated in Fig.~\ref{fig:LCII-SFR}: DEIMOS\_COSMOS\_873756 (hereafter DC873756; also referred to as CRISTAL-24 in \citet{HerreraCamus+25}), DEIMOS\_COSMOS\_818760 (DC818760; CRISTAL-23), and VUDS\_COSMOS\_5110377875 (VC5110377875). The targets were chosen for their expected large molecular gas reservoirs inferred by \citet{Dessauges+20} from the empirical correlation between $L_{\rm [CII]}$ and $M_{\rm molgas}$ \citep{Zanella+18}.

Moreover, as ALPINE galaxies have been shown to span a variety of galaxy types already in place only $1-1.5$~Gyr after the Big Bang \citep{LeFevre+20,Jones+21}, we specifically selected the three most [C\,{\sc ii}]-luminous galaxies exhibiting distinct ISM morpho-kinematic properties, as determined from the ALPINE $0.8''-1.0''$ resolution observations, with the aim of exploring the detectability of mid-$J$ CO lines across different types of high-redshift MS galaxies.
DC873756 at $z=4.5457$ (red hexagon in Figs.~\ref{fig:LCII-SFR} and \ref{fig:MS}), the brightest [C\,{\sc ii}] emitter in the ALPINE sample with $L_{\rm [CII]} = (4.39\pm 0.17)\times 10^9~L_{\odot}$, is a dispersion-dominated galaxy showing no evidence of rotation in the [C\,{\sc ii}] moment-1 map \citep{Jones+21}. Recent ALMA higher-resolution [C\,{\sc ii}] observations (synthesised beam size of $0.32\arcsec\times 0.26\arcsec$; see Fig.~\ref{fig:DC873756}) reveal a bright compact central [C\,{\sc ii}] component surrounded by a fainter extended diffuse halo and two faint [C\,{\sc ii}] clumps \citep{Bethermin+23,Devereaux+24,HerreraCamus+25,Lee+25}. 
DC818760 at $z=4.5613$ (red square in Figs.~\ref{fig:LCII-SFR} and \ref{fig:MS}) is a major merger system \citep{Jones+20,Jones+21,Romano+21} composed of three galaxies, now fully resolved in the new ALMA high-resolution [C\,{\sc ii}] observations (synthesised beam size of $0.30\arcsec\times 0.23\arcsec$; see Fig.~\ref{fig:DC818760}). The eastern and central galaxies are merging and together have a total $L_{\rm [CII]} = (5.74\pm 0.30)\times 10^9~L_{\odot}$ \citep{Devereaux+24,HerreraCamus+25,Lee+25}. 
VC5110377875 at $z=4.5505$ (red star in Figs.~\ref{fig:LCII-SFR} and \ref{fig:MS}) is the most [C\,{\sc ii}]-luminous rotating galaxy in the ALPINE sample with $L_{\rm [CII]} = (1.59\pm 0.11)\times 10^9~L_{\odot}$. It exhibits clear ordered rotation in the [C\,{\sc ii}] moment-1 map, while the velocity dispersion peak in the moment-2 map is slightly offset from the [C\,{\sc ii}] intensity peak \citep{Jones+21}.

The physical properties derived for the three ALPINE targets selected for this study are reported in the following publications:
\begin{itemize}
\item \citet{Bethermin+20} for $L_{\rm [CII]}$ \citep[see also][]{Devereaux+24}, the $158~\mu$m continuum luminosities, and the IR luminosities ($L_{\rm IR}$) integrated between 8~$\mu$m and 1000~$\mu$m\footnote{The rest-frame 158~$\mu$m continuum was detected in VC5110377875 only after combining the original ALPINE data with higher-angular-resolution ALMA follow-up observations \citep{Bethermin+23,Devereaux+24}.};
\item \citet{Faisst+20} for $M_{\rm stars}$ and $\rm SFR_{SED}$ derived from LePhare SED fitting including the HST optical-to-near-IR and {\it Spitzer} IR photometry, and the UV luminosities at 1500~\AA\ rest-frame; and 
\item \citet{Dessauges+20} for $M_{\rm molgas}$ determined from: 
(1)~$L_{\rm [CII]}$ via the \citet{Zanella+18} calibration ($M_{\rm molgas}^{\rm [CII]}$);
(2)~the rest-frame 850~$\mu$m luminosities ($L_{850\mu{\rm m}}$), obtained by extrapolating the rest-frame 158~$\mu$m continuum assuming a modified blackbody (MBB) FIR SED with a cold mass-weighted dust temperature of $T_{\rm dust} = 25$~K and a dust emissivity index of $\beta = 1.8$, and converted into molecular gas masses ($M_{\rm molgas}^{850\mu{\rm m}}$) using the \citet{Scoville+16} calibration, which assumes optically thin dust emission in the Rayleigh-Jeans regime; and
(3)~the [C\,{\sc ii}]-based dynamical masses combined with $M_{\rm stars}$ ($M_{\rm molgas}^{\rm dyn}$). 
\end{itemize}

The dynamical masses ($M_{\rm dyn}$) of DC873756 and VC5110377875 have been derived by \citet{Dessauges+20} from spatially resolved [C\,{\sc ii}] observations, assuming, respectively, a virialised spherical system and a disk-like gas distribution, with the disk inclination estimated from the [C\,{\sc ii}] axis ratio, which has also been used to determine the galaxies' effective radii \citep{Fujimoto+20}. In contrast, no reliable $M_{\rm dyn}$ could be derived for the merger DC818760 because of its complex dynamics. The inferred $M_{\rm dyn}$ were assumed to trace the total baryonic masses, under the assumption of a low dark matter contribution in the inner galaxy regions \citep{Barnabe+12}. More recent studies suggest dark matter fractions of up to $\sim 50$\% in massive galaxies at cosmic noon \citep{Puglisi+23,Nestor+23}, which could reduce $M_{\rm molgas}^{\rm dyn}$ by $\sim 0.1-0.2$~dex of our $z\sim 4.5$ galaxies, but still remaining within measurement uncertainties. 

JWST NIRSpec IFU spectroscopy has recently been acquired for a sub-sample of ALPINE galaxies, including DC873756 and VC5110377875. Using strong nebular emission-line diagnostics and the calibrations of \citet{Sanders+24}, \citet{Faisst+26} derived gas-phase metallicities of $12+\log({\rm O/H}) = 8.68^{+0.58}_{-0.88}$ for DC873756, consistent with a solar metallicity, and $12+\log({\rm O/H}) = 8.41^{+0.08}_{-0.09}$ for VC5110377875, corresponding to a subsolar metallicity. 

\begin{table*}[!]
\caption{Physical properties.}             
\label{tab:properties}      
\centering          
\begin{tabular}{l l l l l} 
\hline\hline \\[-9pt]     
Physical parameter & DC873756 & DC818760~ \tablefootmark{$(\dag)$} & VC5110377875 & References \\
\hline \\[-9pt]
Morpho-kinematic classification & dispersion-dominated & merger & rotator & 1, 2 \\
$z_{\rm [CII]}$ & 4.5457 & 4.5613 & 4.5505 & 3 \\
$L_{\rm [CII]}$ $(L_{\odot})$ & $(4.39\pm 0.17)\times 10^9$  & $(5.74\pm 0.30)\times 10^9$ & $(1.59\pm 0.11)\times 10^9$ & 2, 3 \\
$r_{\rm eff,[CII]}$ (kpc) & $2.36\pm 0.11$ & -- & $2.63\pm 0.23$ & 4 \\
$L_{\rm IR}$ $(L_{\odot})$ & $(1.83\pm 0.10)\times 10^{12}$ & $(1.46\pm 0.18)\times 10^{12}$ & $(0.70\pm 0.16)\times 10^{12}$~ \tablefootmark{$(\ddag)$} & 3 \\
$\rm SFR_{UV+IR}$ $(M_{\odot}~\rm yr^{-1})$~ \tablefootmark{(a)} & $195\pm 72$ & $184\pm 55$ & $106\pm 32$ & 3, 5 \\
$\log (M_{\rm stars})$ $(M_{\odot})$~ \tablefootmark{(b)} & $10.25\pm 0.09$ & $10.63\pm 0.10$ & $10.17\pm 0.16$ & 5 \\
$12+\log ({\rm O/H})$~ \tablefootmark{(c)} & $8.68^{+0.58}_{-0.88} $ & -- & $8.41^{+0.08}_{-0.09}$ & 6 \\
$\log (M_{\rm molgas}^{\rm [CII]})$ $(M_{\odot})$~ \tablefootmark{(d)} & $11.14\pm 0.30$ & $11.26\pm 0.30$ & $10.70\pm 0.30$ & 7 \\
$\log (M_{\rm molgas}^{850\mu{\rm m}})$ $(M_{\odot})$~ \tablefootmark{(e)} & $11.00\pm 0.10$ & $10.90\pm 0.12$ & $10.60\pm 0.13$~ \tablefootmark{$(\ddag)$} & 7 \\
$\log (M_{\rm molgas}^{\rm dyn})$ $(M_{\odot})$~ \tablefootmark{(f)} & $11.23\pm 0.10$ & -- & $10.60\pm 0.18$ & 7 \\
$\log (M_{\rm molgas}^{\rm CO(4-3)})$ $(M_{\odot})$~ \tablefootmark{(g)} & $11.15\pm 0.16$ & $10.62\pm 0.21$ & $<10.69$ & This work \\
$\log (M_{\rm molgas}^{\rm CO(5-4)})$ $(M_{\odot})$~ \tablefootmark{(g)} & $11.02\pm 0.17$ & $<10.80$ & $<10.68$ & This work \\
$\log (M_{\rm molgas}^{\rm [CI](1-0)})$ $(M_{\odot})$~ \tablefootmark{(h)} & $<11.09$ & $<11.12$ & $<10.95$ & This work \\
\hline
\end{tabular}
\tablefoot{References: 1. \citet{Jones+21}; 2. \citet{Devereaux+24}; 3. \citet{Bethermin+20}; 4. \citet{Fujimoto+20}; 5. \citet{Faisst+20}; 6. \citet{Faisst+26}; 7. \citet{Dessauges+20}.
\tablefoottext{$\dag$}{All the listed measurements correspond to the integrated values of the merging eastern and central galaxies; the western galaxy is not included (see Fig.~\ref{fig:DC818760}).}
\tablefoottext{a}{Total star formation rates including both the contribution of the UV luminosity at 1500~\AA\ rest-frame (uncorrected for dust attenuation) and $L_{\rm IR}$.}
\tablefoottext{b}{The stellar masses of \citet{Faisst+20} quoted here have been recently reviewed either by using the CIGALE SED fitting code \citep{Mitsuhashi+24b}, by including JWST/NIRCam photometry \citep{Li+24}, or by performing pixel-by-pixel SED fitting using spatially integrated nebular emission lines \citep{Tsujita+25}. All lead to consistent $M_{\rm stars}$ measurements within $1-2\sigma$ uncertainty.}
\tablefoottext{c}{Gas-phase metallicities derived from the JWST NIRSpec IFU spectroscopy using strong nebular emission line methods and calibrations of \citet{Sanders+24}. There is no JWST NIRSpec spectroscopy for DC818760.}
\tablefoottext{d}{Molecular gas masses derived from $L_{\rm [CII]}$ via the \citet{Zanella+18} relation: $\log(L_{\rm [CII]}) = (-1.28\pm 0.21)+(0.98\pm 0.02)\log(M_{\rm molgas}^{\rm [CII]})$, corresponding to the fiducial [C\,{\sc ii}] to gas mass conversion factor $\alpha_{\rm [CII]}^{\rm Zanella} = M_{\rm molgas}/L_{\rm CII} = 31~M_{\odot}~L_{\odot}^{-1}$.}
\tablefoottext{e}{Molecular gas masses derived using the \citet{Scoville+16} calibration and the rest-frame 850~$\mu$m luminosity extrapolated from the observed rest-frame 158~$\mu$m continuum by assuming the MBB SED with a cold mass-weighted $T_{\rm dust} = 25$~K and $\beta = 1.8$.}
\tablefoottext{f}{Molecular gas masses derived from $M_{\rm dyn}$ and $M_{\rm stars}$ by assuming negligible dark-matter contribution in the internal regions of galaxies: $M_{\rm molgas}^{\rm dyn} = M_{\rm dyn} - M_{\rm stars}$.}
\tablefoottext{g}{Molecular gas masses derived from the CO(4--3) and CO(5--4) luminosities using Eq.~\ref{eq:MmolgasCO} and assuming $\alpha_{\rm CO}^{\rm MW} = 4.36~M_{\odot}~({\rm K~km~s^{-1}~pc^2})^{-1}$ and the \citet{Boogaard+20} $r_{4,1} = 0.61\pm 0.13$ and $r_{5,1} = 0.44\pm 0.11$ derived for MS galaxies at $z = 2-2.7$.}
\tablefoottext{h}{Molecular gas masses derived from the [C\,{\sc i}](1--0) luminosities using Eq.~\ref{eq:MmolgasCI} and assuming $X_{\rm [CI]} = (1.41\pm 0.07)\times 10^{-5}$ and $Q_{10} = 0.48\pm 0.08$ from \citet{Dunne+22}.}
\tablefoottext{$\ddag$}{Derived from the rest-frame 158~$\mu$m continuum detection obtained when combining the low- and high-resolution ALMA observations \citep{Bethermin+23,Devereaux+24}.}
All upper limits are obtained from the $3\sigma$ luminosity upper limits listed in Table~\ref{tab:observations}.
}
\end{table*}
\begingroup
\setlength{\tabcolsep}{2.8pt} 
\begin{table*}[!]
\caption{NOEMA observations.}             
\label{tab:observations}      
\centering          
\begin{tabular}{l c c c c c c c c} 
\hline\hline \\[-9pt]     
Target & Line & $\nu_{\rm obs}$ & Sideband & Synthesised beam & RMS~ \tablefootmark{(a)} & FWHM~ \tablefootmark{(b)} & $I_{\rm line}$~ \tablefootmark{(c)} & $L^{\prime}_{\rm line}$~ \tablefootmark{(d)} \\
 & & (GHz) & & Size ($\arcsec$)/PA($^{\circ}$) & ($\rm mJy~beam^{-1}$) & ($\rm km~s^{-1}$) & ($\rm mJy~km~s^{-1}$) & ($10^{10}~\rm K~km~s^{-1}~pc^2$) \\
\hline \\[-9pt]     
DC873756 & CO(4--3)  & 83.13 & LSB & $12.7\times 2.9$/$+9\phantom{1}$ & 0.15 & $351\pm 25$ & $395\pm 87$ & $1.97\pm 0.43$ \\
                  & CO(5--4)  & 103.91 & USB & $11.5\times 2.4$/$+9\phantom{1}$ & 0.16 & $330\pm 26$ & $324\pm 71$ & $1.04\pm 0.23$ \\
                  & [C\,{\sc i}](1--0) & 88.74 & LSB & $12.7\times 2.9$/$+9\phantom{1}$ & 0.14 & -- & $<148$ & $<0.65$ \\[4pt]  
DC818760 & CO(4--3)  & 82.90 & LSB & $\phantom{1}5.7\times 2.2$/$+13$ & 0.20 & $121\pm 9$ & $115\pm 44$~ \tablefootmark{($\dag$)} & $0.58\pm 0.22$~ \tablefootmark{($\dag$)} \\
                  & CO(5--4)  & 103.62 & USB & $\phantom{1}4.3\times 1.8$/$+15$ & 0.18 & -- & $<198$ & $<0.64$ \\
                 & [C\,{\sc i}](1--0) & 88.50 & LSB & $\phantom{1}5.7\times 2.2$/$+13$ & 0.15 & -- & $<159$ & $<0.70$ \\[4pt]  
VC5110377875 & CO(4--3)  & 83.06 & LSB & $\phantom{1}6.0\times 3.2$/$+15$ & 0.12 & -- & $<137$ & $<0.68$ \\
                         & CO(5--4)  & 103.82 & USB & $\phantom{1}5.0\times 2.6$/$+16$ & 0.14 & -- & $<154$ & $<0.49$ \\
                         & [C\,{\sc i}](1--0) & 88.67 & LSB & $\phantom{1}6.0\times 3.2$/$+15$ & 0.10  & -- & $<106$ & $<0.47$ \\
\hline
\end{tabular}
\tablefoot{
\tablefoottext{a}{RMS noise levels per 20~MHz channel and per beam of the cleaned spectral line data cubes at the frequencies of the respective lines and at the phase center.}
\tablefoottext{b}{Full width half maximum derived from the Gaussian best fit to the observed CO line profile when detected (Fig.~\ref{fig:spectra}).}
\tablefoottext{c}{The velocity-integrated line fluxes include a correction for the CMB radiation of 21\% for CO(4--3), 18\% for CO(5--4), and 15\% for [C\,{\sc i}](1--0) (see Sect.~\ref{sect:COCI-luminosities}). For non-detections, we provide $3\sigma$ upper limits assuming a FWHM of $300~\rm km~s^{-1}$ for both the CO and [C\,{\sc i}] lines, motivated by the FWHM of the CO lines detected in DC873756 and the [C\,{\sc ii}] lines observed in the ALPINE galaxy sample \citep{Bethermin+20}.}
\tablefoottext{d}{Luminosities derived from $I_{\rm line}$ using Eq.~\ref{eq:L'}.}
\tablefoottext{$\dag$}{Measurements derived based on the tentative $4.3\sigma$ detection of the CO(4--3) emission line.}
}
\end{table*}
\endgroup

In Table~\ref{tab:properties} we summarise the physical properties of the three galaxies. These [C\,{\sc ii}]-luminous systems lie at the low redshift ($z\sim 4.5$), massive ($M_{\rm stars} >10^{10}~M_{\odot}$), and highly star-forming end of the ALPINE galaxy distribution, with DC873756 and DC818760 having among the higher SFRs (${\rm SFR_{UV+IR}}\sim 190~M_{\odot}~\rm yr^{-1}$). 
As also shown in Fig.~\ref{fig:MS}, although they exhibit distinct morpho-kinematic properties, the three galaxies are representative of massive, highly star-forming MS galaxies at $z\sim 5$ (see the grey shaded band).


\section{NOEMA observations and data reduction}
\label{sect:observations}

The three ALPINE galaxies were observed with NOEMA under the program ID W22DW (PIs: M.~Dessauges-Zavadsky \& M.~B\'ethermin) in winter 2023. For each target the frequency tuning of the 31~GHz instantaneous bandwidth of the PolyFiX correlator was tailored in band~1 to simultaneously sample the CO(4--3), CO(5--4), and [C\,{\sc i}](1--0) 609~$\mu$m emission lines with a fixed channel spacing of 2~MHz resolution in dual polarisation. Array configurations C or D (see the resulting synthesised beam sizes after imaging in Table~\ref{tab:observations}) were used to maximise the sensitivity and minimise flux losses from resolved emission. While the large NOEMA beam may reduce the observed surface brightness through beam dilution, it does not affect the integrated flux measurements used throughout our analysis. The requested NOEMA sensitivities were estimated from the expected CO fluxes inferred from $M_{\rm molgas}^{\rm [CII]}$ (Table~\ref{tab:properties}). 
To achieve these fluxes, the total observing time per target was set to 13~hours, 13~hours, and 23~hours for DC873756, DC818760, and VC5110377875, respectively.

\begin{figure*}[!]
\centering
\includegraphics[width=0.3\textwidth,clip]{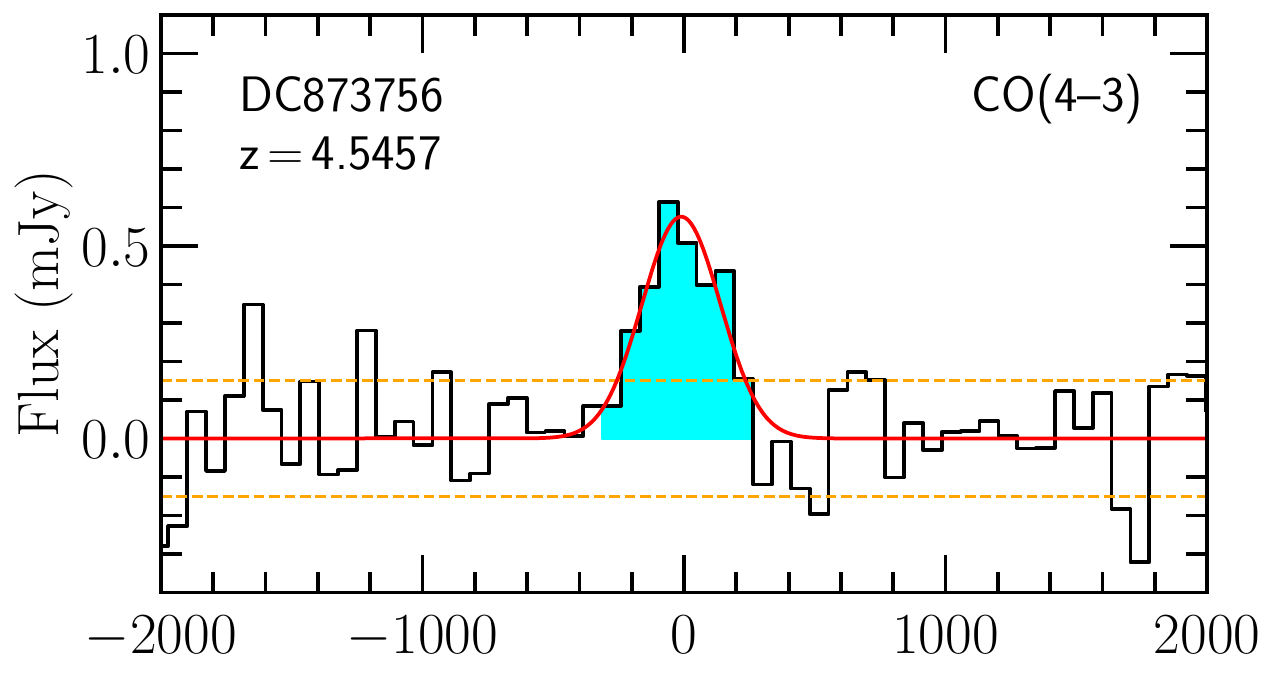}
\includegraphics[width=0.3\textwidth,clip]{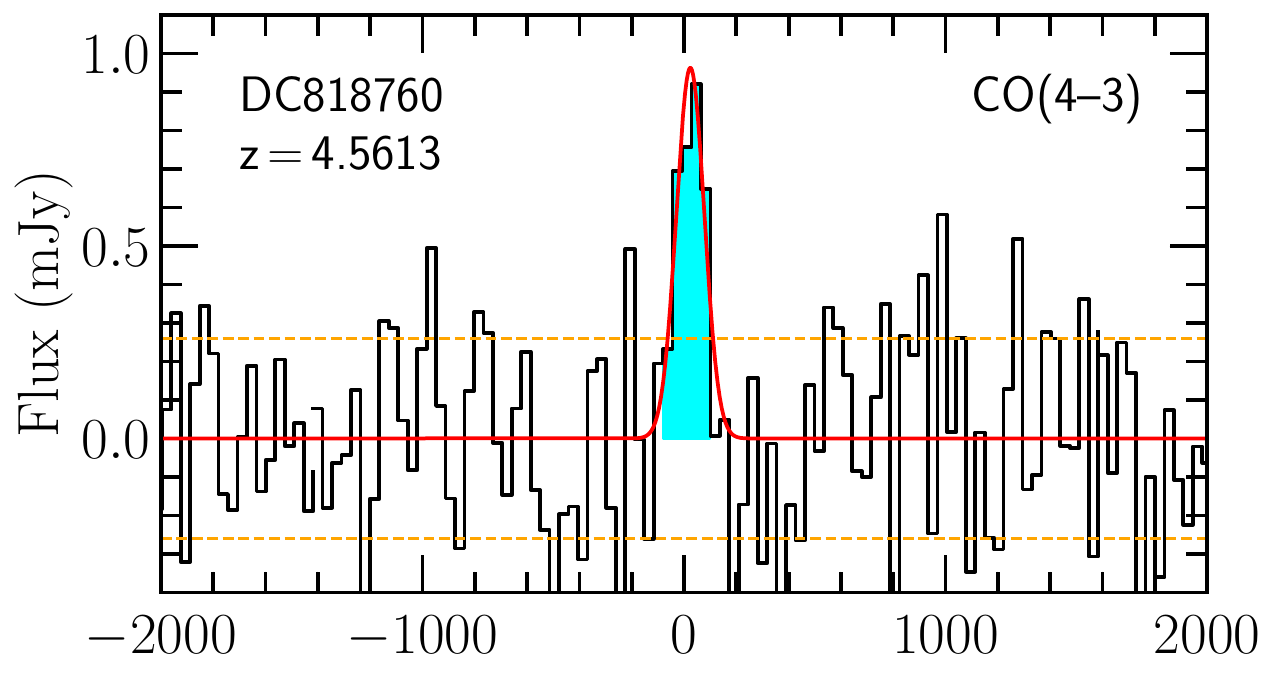}
\vspace{-0.4cm}
\includegraphics[width=0.3\textwidth,clip]{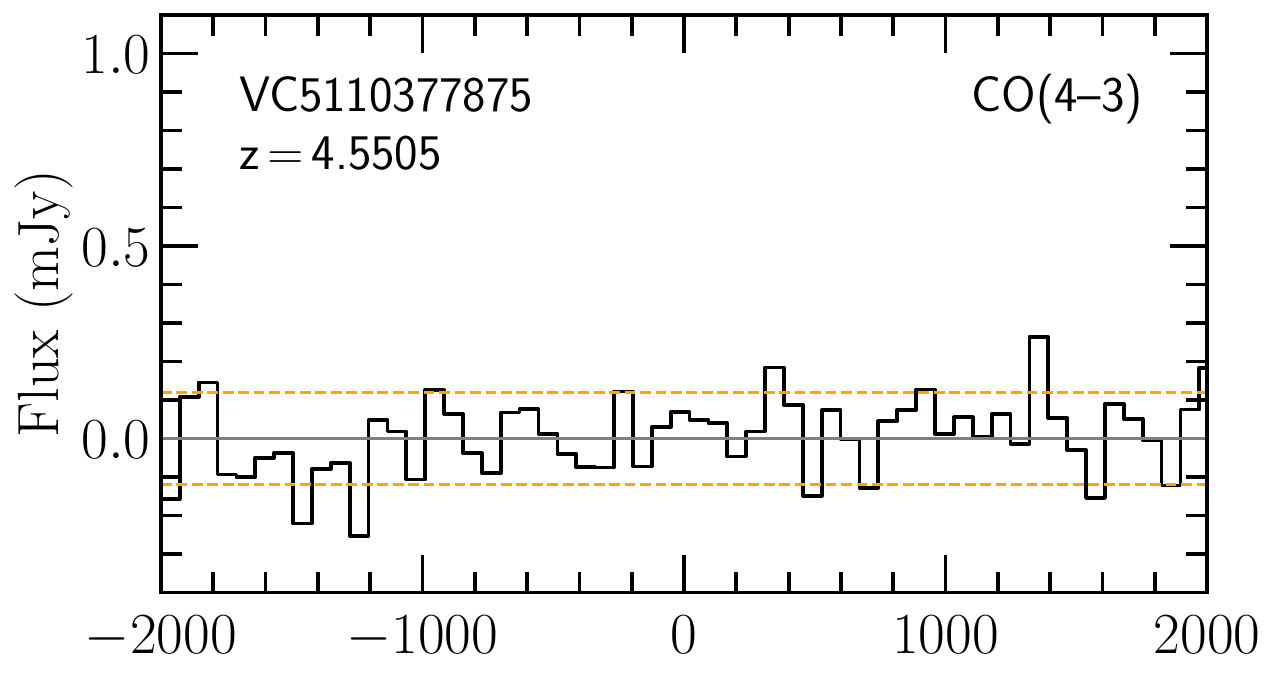}
\includegraphics[width=0.3\textwidth,clip]{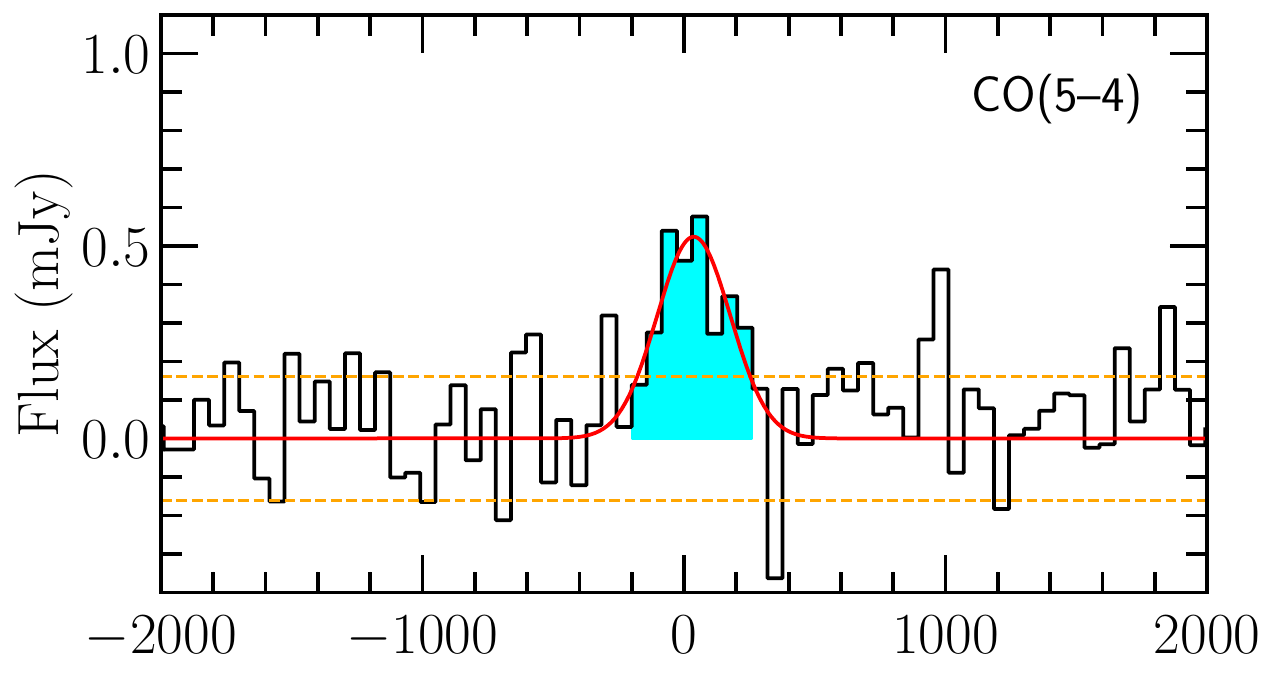}
\includegraphics[width=0.3\textwidth,clip]{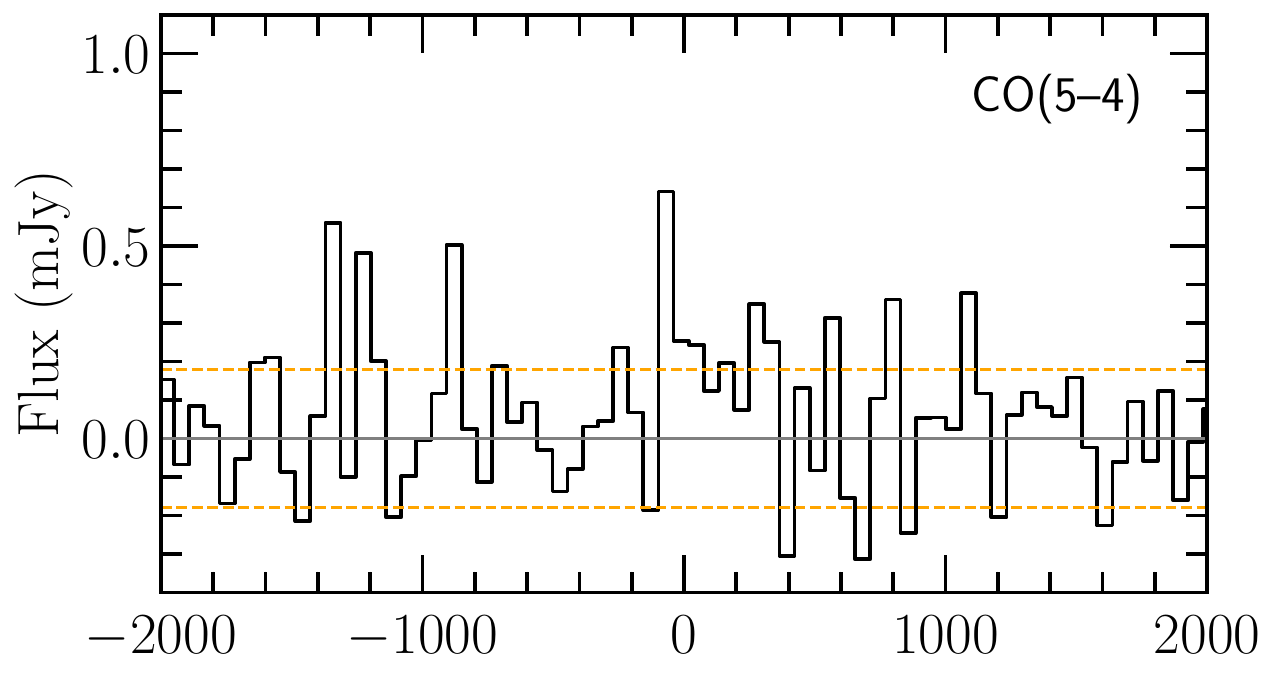}
\vspace{-0.4cm}
\includegraphics[width=0.3\textwidth,clip]{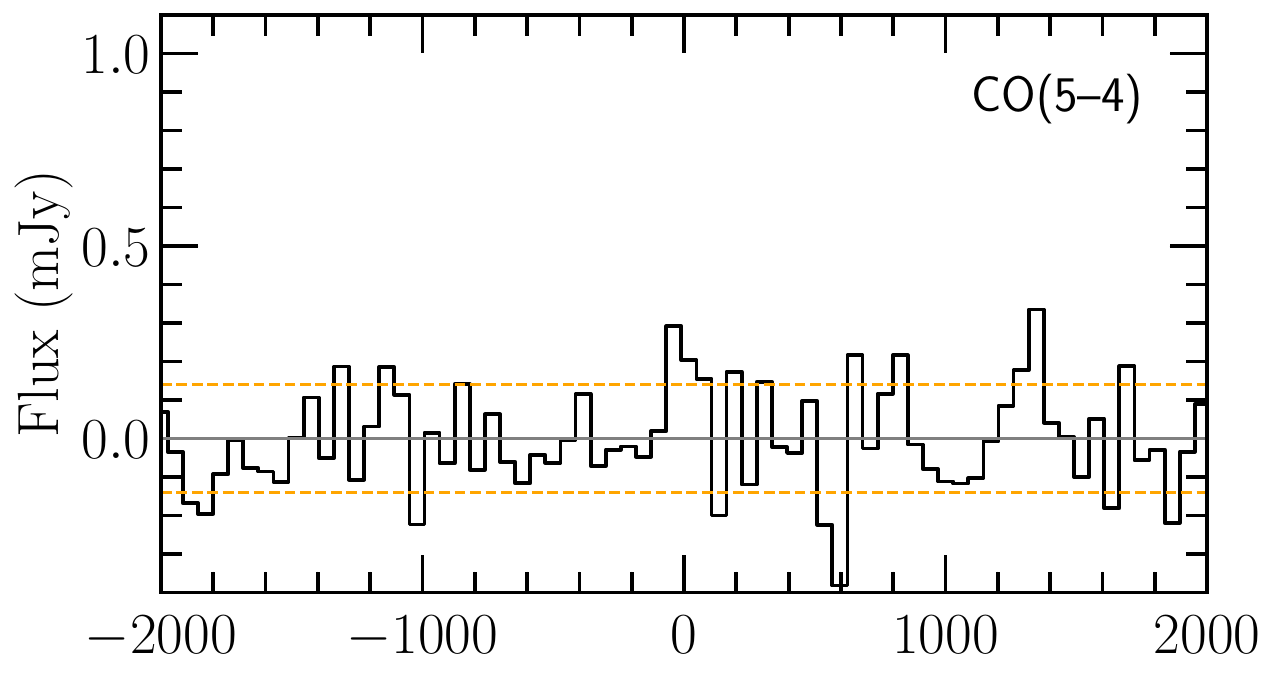}
\includegraphics[width=0.3\textwidth,clip]{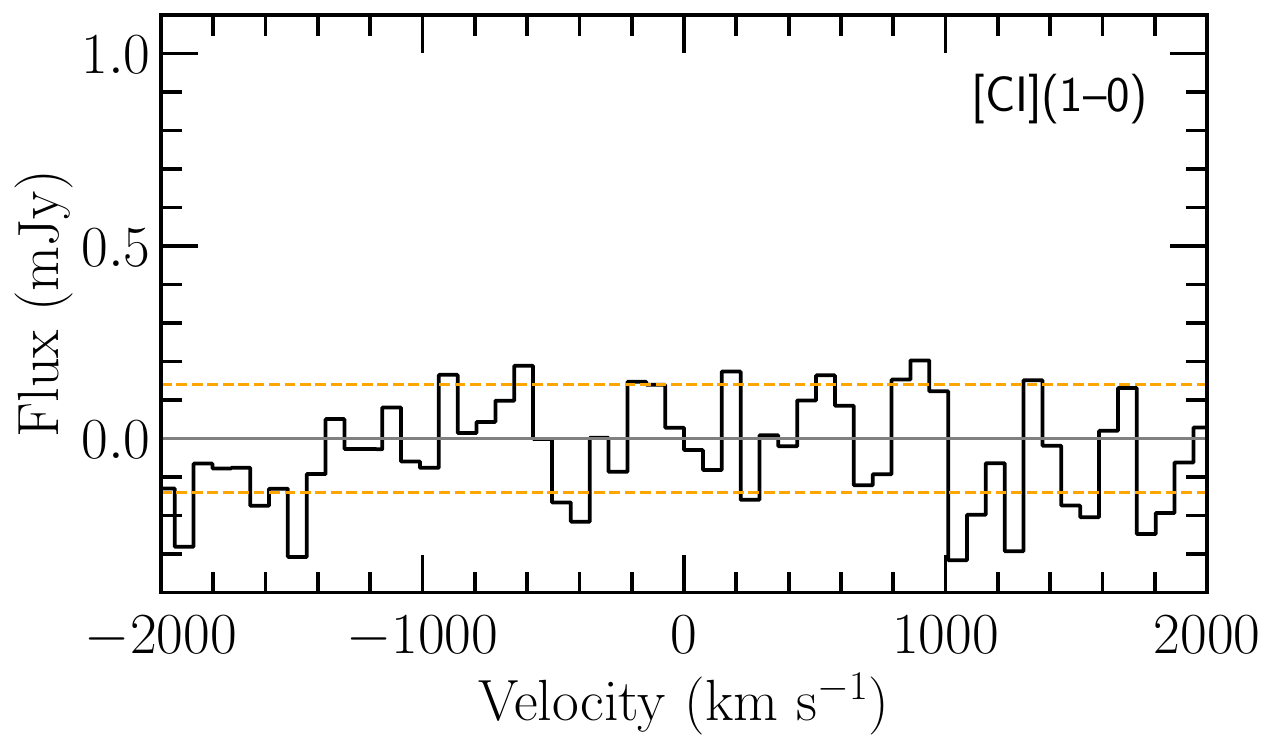}
\includegraphics[width=0.3\textwidth,clip]{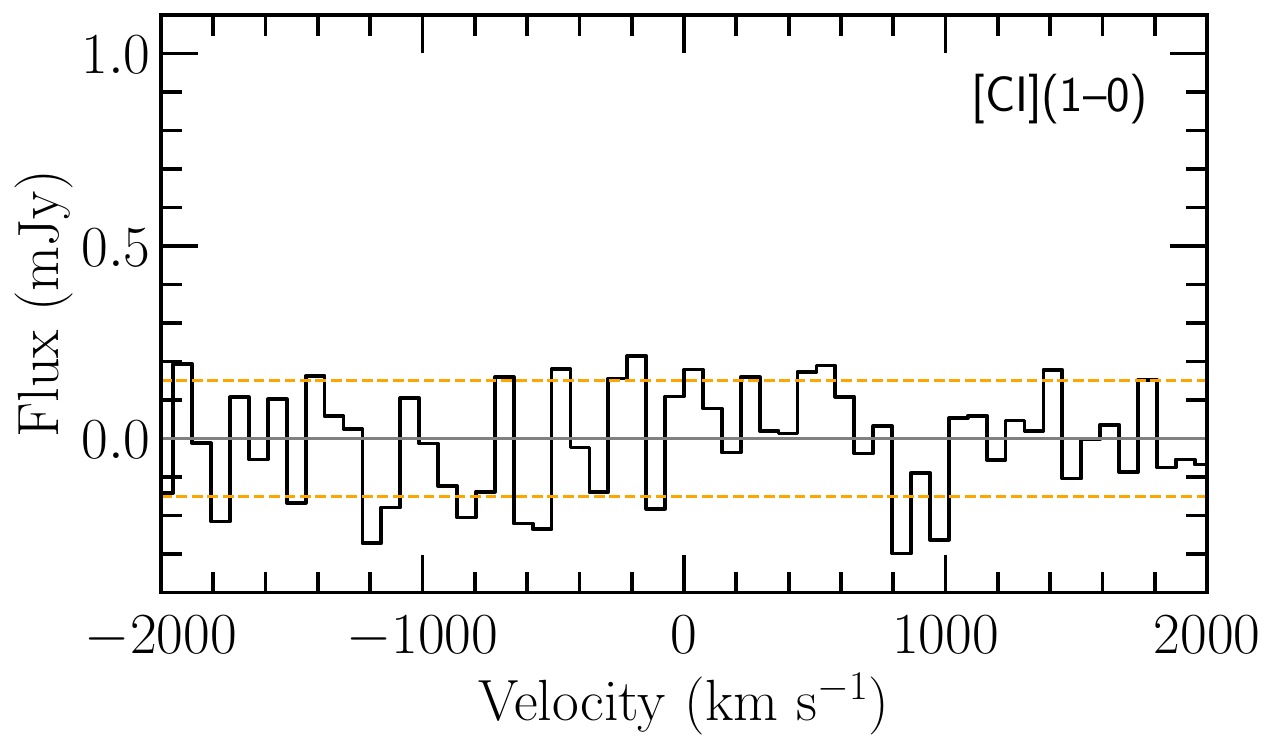}
\includegraphics[width=0.3\textwidth,clip]{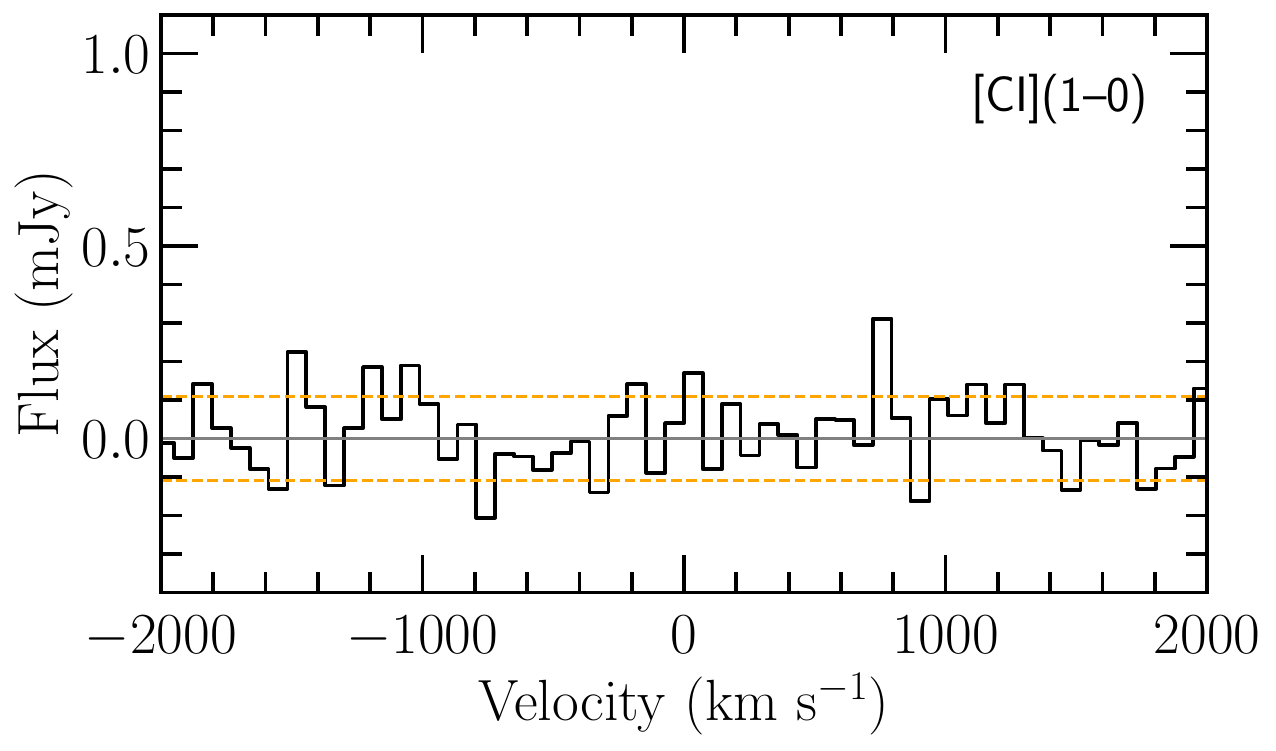}
\caption{NOEMA CO(4--3), CO(5--4), and [C\,{\sc i}](1--0) emission line spectra acquired for three [C\,{\sc ii]}-luminous ALPINE galaxies, DC873756, DC818760, and VC5110377875, plotted from top to bottom in steps of 20~MHz (except for CO(4--3) in DC818760 plotted in steps of 10~MHz) and with the zero velocity centred on the redshift derived from the [C\,{\sc ii}] emission line \citep{Bethermin+20}. The dashed orange lines correspond to the RMS noise level of spectra averaged over spectral channels in the range from $-2000~\rm km~s^{-1}$ to $+2000~\rm km~s^{-1}$ around the expected frequency of the line. 
The solid red lines are the Gaussian best-fits to the observed line profiles when detected.}
\label{fig:spectra}
\end{figure*}

The NOEMA data were calibrated with the support of IRAM astronomers using the GILDAS\footnote{\url{https://www.iram.fr/IRAMFR/GILDAS/}} software package \texttt{CLIC} with the flux, bandpass, and phase calibration performed with the calibrators most suitable for our targets. The resulting  calibrated visibilities ($uv$ tables) were then imaged using the Common Astronomy Software Application \citep[CASA, version 6.2.1.7;][]{McMullin+07} applying a pixel size of $0.4\arcsec$, a spectral channel width of 10~MHz or 20~MHz, and the natural weighting to maximise sensitivity. The clean used to remove side lobes was performed using the \texttt{tclean} routine and was repeated down to the threshold of 4 times the RMS noise level of the dirty spectral line cubes. The CO(4--3) and CO(5--4) emission lines were detected in DC873756 at high confidence levels
of $10.5\sigma$ and $8.6\sigma$, respectively, while only a tentative $4.3\sigma$ detection of the CO(4--3) line was obtained for DC818760, and no CO line was detected in VC5110377875 (see Figs. A.1, A.2, and \ref{fig:spectra}). We obtained no detection of the [C\,{\sc i}](1--0)  line. We also imaged the band~1 calibrated visibilities of continuum, excluding channels contaminated by the emission lines when detected, using the same pixel size of $0.4\arcsec$ and the natural weighting. No 3~mm ($\sim 545~\mu \rm m$ rest-frame) dust continuum was detected. The resulting synthesised beam sizes and RMS noise levels at the phase center of the spectral line cubes for the three targets are listed in Table~\ref{tab:observations}.

\section{Analysis and results}
\label{sect:results}

\subsection{CO and [C\,{\sc i}] flux and luminosity measurements}
\label{sect:COCI-luminosities}

To derive the velocity-integrated fluxes of the CO(4--3) and CO(5--4) emission lines detected in DC873756, we followed the method of \citet{Novak+20} optimised for unresolved sources. We first extracted the 1D spectra of the respective CO emission lines by iteratively testing different spectral channel ranges to determine the one that maximises the signal-to-noise ratio \citep[see e.g.][]{Daddi+15,Zanella+18}. We then performed the Gaussian fitting of the CO lines using the nonlinear $\chi^2$ minimisation and the Levenberg-Marquardt algorithm to derive the full-width half maximum (FWHM). We averaged the spectral channels over $1.2\times \rm FWHM$ \citep[which accounts for 84\% of the line flux;][]{Novak+20} to produce the CO velocity-integrated moment-0 maps employing the \texttt{immoments} routine in CASA. We measured the respective peak fluxes in the moment-0 maps and inferred from them the total CO(4--3) and CO(5--4) velocity-integrated line fluxes ($I_{\rm CO(4-3)}$ and $I_{\rm CO(5-4)}$). 

We used the same method for the tentative $4.3\sigma$ CO(4--3) line detection in DC818760. 
We consider this CO(4--3) line detection as tentative because of both its low significance level and its small FWHM of $126\pm 9~\rm km~s^{-1}$ in comparison to the [C\,{\sc ii}] FWHM of $275~\rm km~s^{-1}$ \citep{Bethermin+20}.

For the CO and [C\,{\sc i}] non-detections, $3\sigma$ upper limits on the respective velocity-integrated line fluxes were derived from the RMS noise level of the cleaned spectral line data cubes at the frequencies of the respective lines and at the phase center. A FWHM of $\sim 300~\rm km~s^{-1}$ was assumed for both the CO and [C\,{\sc i}] lines given the CO(4--3) and CO(5--4) FWHM in DC873756 (Table~\ref{tab:observations}) and the mean FWHM of $280~\rm km~s^{-1}$ of [C\,{\sc ii}] lines measured for ALPINE galaxies \citep{Bethermin+20}. 

We used Eq.~3 from \citet{Solomon+97} to derive the CO and [C\,{\sc i}] luminosities in $\rm K~km~s^{-1}~pc^2$:
\begin{equation}
\label{eq:L'}
L^{\prime}_{\rm line} = 3.25\times 10^7 I_{\rm line} \nu_{\rm obs}^{-2} (1+z)^{-3} D_{\rm L}^2,
\end{equation}
where $I_{\rm line}$ is the velocity-integrated line flux in $\rm Jy~km~s^{-1}$, $\nu_{\rm obs}$ is the observed line frequency in GHz, $z$ is the redshift of the galaxy, and $D_{\rm L}$ is the luminosity distance in Mpc. As the CMB temperature increases with redshift, reaching $\sim 15$~K at $z\sim 4.5$, the redshift of the studied galaxies, the effect of the CMB radiation on the $I_{\rm line}$ measurements must also be taken into account.

While the effect of the CMB on $I_{\rm [CII]}$ is expected to be weak over the ALPINE redshift range, as discussed by \citet{Bethermin+20} based on the studies of \citet{Vallini+15} and \citet{Lagache+18}, this is not the case for the CO(4--3) and CO(5--4) transitions. \citet{daCunha+13} computed the ratios between the velocity-integrated line fluxes observed against the CMB and the intrinsic line fluxes for different CO transitions and redshifts, finding that the CMB can significantly reduce the observed CO fluxes at high redshifts in both the local thermal equilibrium (LTE) and non-LTE cases (see also \citealt{Zhang+16} for the effects on low-$J$ CO transitions). Assuming LTE conditions and a gas kinetic temperature of $T_{\rm kin} = 40$~K for the three studied galaxies, comparable to the mean dust temperature of $T_{\rm dust} = 47\pm 2$~K derived from the stacked SED of ALPINE analogues at $4<z<5$ with ${\rm SFR} > 10~M_{\odot}~\rm yr^{-1}$ \citep{Bethermin+20}, the expected line flux ratios are $\sim 0.79$ for CO(4--3) and $\sim 0.82$ for CO(5--4) at $z\sim 4.5$ (see Fig.~6 of \citealt{daCunha+13}). Comparable CO(4--3) and CO(5--4) line flux ratios are obtained in the non-LTE case for the same $T_{\rm kin}$ and a high density of $n_{\rm H_2}\sim 10^4~\rm cm^{-3}$. This implies corrections of $\sim21$\% and $\sim18$\%, respectively, to the measured $I_{\rm CO(4-3)}$ and $I_{\rm CO(5-4)}$ fluxes of our three ALPINE galaxies.

\citet{Frias+25} have estimated the effect of the CMB on the observed velocity-integrated flux of the [C\,{\sc i}](1--0) line also for different redshifts in the non-LTE case. They have found a correction of $\sim 15$\% at $z\sim 4.5$ for $T_{\rm kin} = 50~\rm K$ and $n_{\rm H_2} = 10^4~\rm cm^{-3}$ (see their Fig.~6 showing a very weak dependence on $n_{\rm H_2}$ over the range of $10^3-10^5~\rm cm^{-3}$), in excellent agreement with the correction derived by \citet{Zhang+16}. Assuming again $T_{\rm kin}\sim T_{\rm dust}$, we apply this correction to the measured $I_{\rm [CI](1-0)}$ upper limits.

The resulting CO and [C\,{\sc i}] velocity-integrated line fluxes corrected for the CMB radiation and the corresponding luminosities are listed in Table~\ref{tab:observations}. The respective line spectra with Gaussian fits (when available) are shown in Fig.~\ref{fig:spectra}, and the [C\,{\sc ii}] moment-0 maps together with the detected CO contours can be found in Appendix~A for DC873756 and DC818760. 


\subsection{From the CO luminosities to molecular gas masses}
\label{sect:COanalysis}

The molecular gas mass (in $M_{\odot}$) can be derived from the luminosity of a CO transition ($L^{\prime}_{{\rm CO}~J\rightarrow J-1}$ in $\rm K~km~s^{-1}~pc^2$) following:
\begin{equation}
\label{eq:MmolgasCO}
M_{\rm molgas}^{\rm CO} = \alpha_{\rm CO} \frac{L^{\prime}_{{\rm CO}~J\rightarrow J-1}}{r_{J,1}}~~{\rm with}~r_{J,1} = \frac{L^{\prime}_{{\rm CO}~J\rightarrow J-1}}{L^{\prime}_{\rm CO(1-0)}}.
\end{equation}
The main difficulties lie first in determining the CO SLED required to recover the luminosity of the fundamental CO(1--0) transition from the observed CO($J\rightarrow J-1$) transition through the CO excitation ratio ($r_{J,1}$), and second in knowing the CO-to-H$_2$ conversion factor ($\alpha_{\rm CO}$ in $M_{\odot}~\rm (K~km~s^{-1}~pc^2)^{-1}$) linking $L^{\prime}_{\rm CO(1-0)}$ to the H$_2$ mass, which both depend on the ISM physical conditions of the galaxy (metallicity, temperature, density, dynamical state).

The CO SLED remains so far unconstrained for MS galaxies at $z>4$. Existing measurements for MS galaxies at cosmic noon suggest more excited CO SLEDs than in local systems, with excitation increasing toward higher redshift \citep[e.g.][]{Daddi+15,Boogaard+20,Dessauges+17}. Stacked CO-flux-limited star-forming galaxies from the ALMA SPECtroscopic Survey \citep[ASPECS;][]{Boogaard+20} at $z=2-2.7$ and a lensed MS galaxy at $z\sim3.6$ yield similar values of $r_{4,1} \sim 0.60$ and $r_{5,1} \sim 0.45$ \citep{Boogaard+20,Dessauges+17}. Comparable CO excitation ratios have also been reported for SMGs and DSFGs at $z\sim2-5$, although with a large source-to-source scatter \citep{Bothwell+13,Spilker+14,Frias+23}. \citet{Frias+23} have searched for correlations between CO excitation and redshift, SFR, SFR surface density ($\Sigma_{\rm SFR}$), 870~$\mu$m flux, offset from the MS, and star formation efficiency. They have found only a weak dependence of $r_{4,1}$ on $\Sigma_{\rm SFR}$ \citep[see also][]{Boogaard+20}, consistent with higher $\Sigma_{\rm SFR}$ driving stronger ambient UV radiation fields and enhanced CO excitation, in agreement with the predictions of \citet{NarayananKrumholz14}. Following these studies, we adopt the average $r_{4,1} = 0.61\pm 0.13$ and $r_{5,1} = 0.44\pm 0.11$ derived by \citet{Boogaard+20} for MS galaxies at $\langle z\rangle=2.5$, as these currently are the higher redshift CO SLED measurements available for MS galaxies and therefore the closer analogues to our $z\sim 4.5$ MS galaxies, while keeping in mind the large uncertainties associated with this assumption.

\begin{figure*}[t]
\centering
\includegraphics[width=0.34\textwidth,clip]{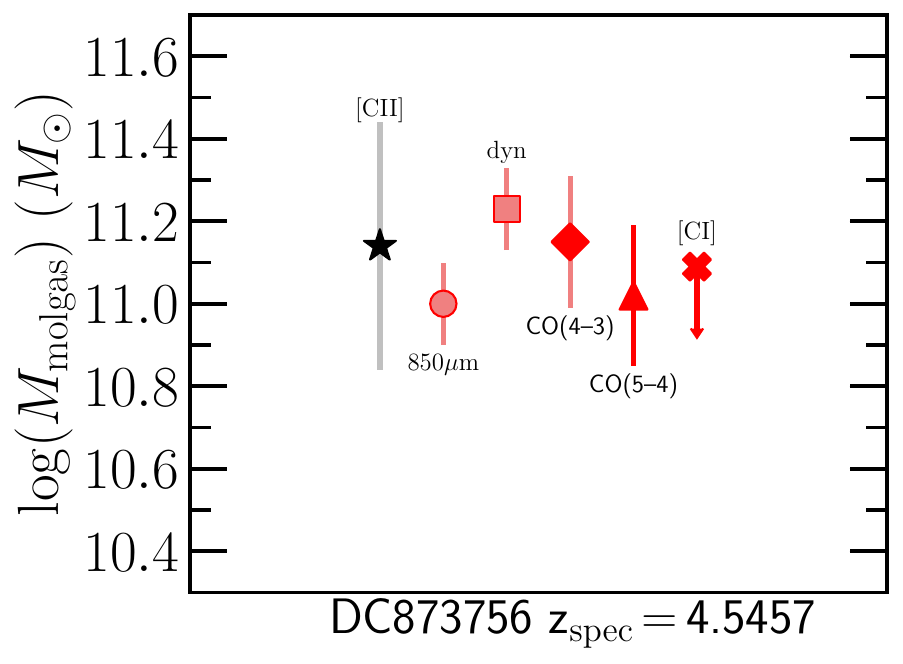}
\includegraphics[width=0.315\textwidth,clip]{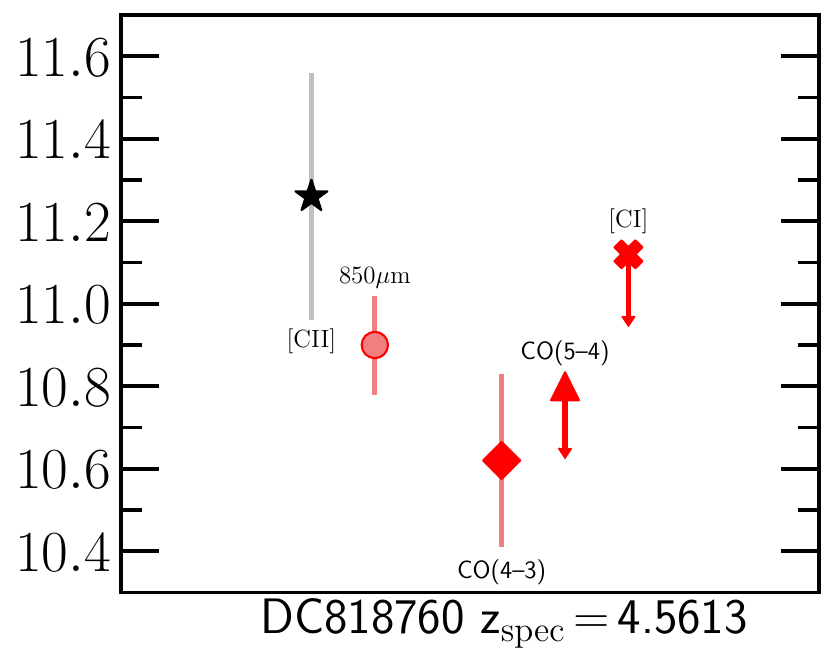}
\includegraphics[width=0.315\textwidth,clip]{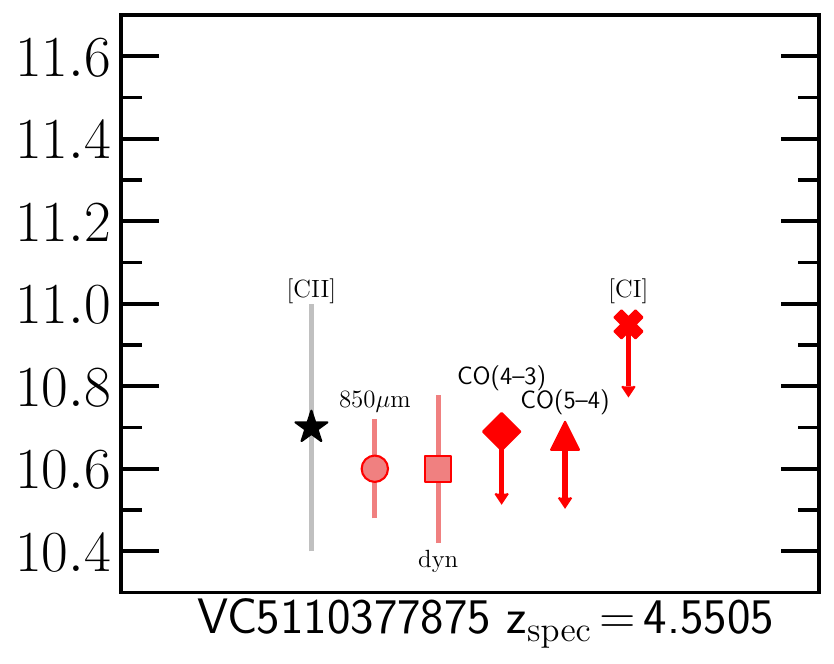}
\caption{Comparison of the molecular gas masses derived from different molecular gas tracers as tabulated in Table~\ref{tab:properties} for the three studied galaxies, DC873756, DC818760, and VC5110377875 from left to right.}
\label{fig:Mmolgas-comparison}
\end{figure*}

The  CO-to-H$_2$ conversion factor has been observationally constrained in the Milky Way and nearby disk galaxies yielding $\alpha_{\rm CO}^{\rm MW} = 4.36~M_{\odot}~({\rm K~km~s^{-1}~pc^2})^{-1}$, while starbursts and (ultra-)luminous IR galaxies ((U)LIRGs) show much lower values of $\alpha_{\rm CO}^{\rm SB} = 0.8-1~M_{\odot}~({\rm K~km~s^{-1}~pc^2})^{-1}$ \citep[both including the factor of 1.36 to account for heavy elements, primarily helium; see the review of][and references therein]{Bolatto+13}. The limited measurements available for MS galaxies and SMGs at cosmic noon favour a similar bimodal distribution, with $\alpha_{\rm CO}^{\rm MW}$ for MS galaxies and $\alpha_{\rm CO}^{\rm SB}$ for SMGs and DSFGs \citep{Dessauges+19,Dessauges+23, Calistro+18,Dunne+22,Amvrosiadis+25}. In addition, a strong metallicity dependence of $\alpha_{\rm CO}$ is now widely supported, with $\log(\alpha_{\rm CO})$ increasing by $\sim0.2$~dex for every 0.1~dex decrease in $\log(Z/Z_{\odot})$ over the range $Z\sim(0.2-2)~Z_{\odot}$ \citep{Genzel+15,Accurso+17,Bisbas+24}. This trend is driven by the reduced production of carbon and oxygen and by the larger fraction of photo-dissociated CO at lower metallicity due to the lower dust abundance available to shield CO from UV radiation fields, leading to increasing amounts of CO-dark molecular gas \citep{Wolfire+10,Leroy+11,Sandstrom+13,Bolatto+13,Sternberg+14,Genzel+15, Madden+20}. Given the MS nature of our galaxies at $z\sim 4.5$, we adopt the conservative $\alpha_{\rm CO}^{\rm MW}$.

The resulting $M_{\rm molgas}^{\rm CO(4-3)}$ and $M_{\rm molgas}^{\rm CO(5-4)}$ of the three ALPINE galaxies at $z\sim 4.5$ are listed in Table~\ref{tab:properties} and shown in Fig.~\ref{fig:Mmolgas-comparison}. 


\subsection{From the [C\,{\sc i}] luminosities to molecular gas masses}
\label{sect:CIanalysis}

Atomic carbon was first proposed as a tracer of molecular gas by \citet{Papadopoulos+04}. Since then, the [C\,{\sc i}](1--0) line has been shown to be optically thin for the bulk of the H$_2$ gas \citep{Weiss+03,Perez+15}, allowing it to probe higher column densities of cold molecular gas than $^{12}$CO lines. It arises from the same regions as CO \citep{Bisbas+15,Glover+15}, but with the advantage of remaining excited in gas as cold as $\sim 24~\rm K$, unlike the [C\,{\sc ii}] line and the mid-$J$ CO transitions that cannot trace cold and subthermally excited gas \citep[$\lesssim  91~\rm K$ and $\lesssim 55~\rm K$ at densities $\lesssim10^3-10^4~\rm cm^{-3}$, respectively;][]{CarilliWalter13}, particularly in the presence of strong UV radiation fields.

The fact that the [C\,{\sc i}](1--0) line is optically thin implies that 
the [C\,{\sc i}] mass can be directly inferred from the [C\,{\sc i}](1--0) luminosity. 
The H$_2$ gas mass can therefore be expressed as a function of $L^{\prime}_{\rm [CI](1-0)}$ in $\rm K~km~s^{-1}~pc^2$, the carbon abundance relative to H$_2$ ($X_{\rm [CI]}$), and the excitation term ($Q_{10}$):
\begin{equation}
\label{eq:MmolgasCI}
M_{\rm molgas}^{\rm [CI]} = 
19.1 \left( \frac{X_{\rm [CI]}}{1.41\times 10^{-5}}\right) ^{-1} \left( \frac{Q_{10}}{0.48}\right) ^{-1} L^{\prime}_{\rm [CI](1-0)},
\end{equation}
where He is included in the factor 19.1 given here in units of $\rm M_{\odot}~(K~km~s^{-1}~pc^2)^{-1}$. $Q_{10}$ describes the relative fraction of atomic carbon in the $J=1$ state. It is a complex function of density and temperature under non-LTE conditions \citep{Papadopoulos+04,Papadopoulos+22,Dunne+22} and has been solved to $Q_{10} = 0.48\pm 0.08$ by \citet{Papadopoulos+22}. $X_{\rm [CI]}$ has been found to vary from $(1-2)\times 10^{-5}$ for MS galaxies at cosmic noon \citep[e.g.][]{Valentino+18,Jiao+19,Boogaard+20,Dunne+22} to $(4-8)\times 10^{-5}$ for ULIRGs and SMGs \citep[e.g.][]{Dunne+22,Gururajan+23,Frias+25}, and to depend linearly on metallicity \citep{Walter+11,HeintzWatson20}. Here we adopt $X_{\rm [CI]} = (1.41\pm 0.07)\times 10^{-5}$ obtained by \citet[][their Table~8]{Dunne+22} for MS galaxies at $z\sim 1$. 

The corresponding $M_{\rm molgas}^{\rm [CI]}$ obtained for the three ALPINE galaxies at $z\sim 4.5$ are given in Table~\ref{tab:properties} and shown in Fig.~\ref{fig:Mmolgas-comparison}. 


\section{Discussion}
\label{sect:discussion}

\subsection{Comparison of gas masses from multi-H$_2$ tracers}
\label{sect:Mmolgas-comparison}

Determining the molecular gas masses of galaxies, particularly at high-redshift, remains challenging because these measurements rely on indirect tracers of the H$_2$ gas. This work presents one of the rare comprehensive comparisons of molecular gas masses derived from up to six different tracers for the same MS galaxies at redshifts as high as $z\sim 4.5$, namely the [C\,{\sc ii}] 158~$\mu$m emission line, the underlying dust continuum, the [C\,{\sc ii}]-based dynamical mass, the CO $J=4$ and $J=5$ emission lines, and the [C\,{\sc i}](1--0) 609~$\mu$m emission line. The respective $M_{\rm molgas}$ measurements and upper limits, obtained under the assumptions described in Sects.~\ref{sect:targets}, \ref{sect:COanalysis}, and \ref{sect:CIanalysis}, are listed in Table~\ref{tab:properties} and compared individually for each galaxy in Fig.~\ref{fig:Mmolgas-comparison}. We discuss the results for each galaxy below.


\subsubsection{DC873756}
\label{sect:DC873756}

The detections of both the CO(4--3) and CO(5--4) emission lines in this MS galaxy at $z = 4.5457$ (Figs.~\ref{fig:spectra} and \ref{fig:DC873756}) allow a unique comparison between the molecular gas masses derived from five different and commonly used tracers: $M_{\rm molgas}^{\rm CO(4-3)}$, $M_{\rm molgas}^{\rm CO(5-4)}$, $M_{\rm molgas}^{\rm [CII]}$, $M_{\rm molgas}^{850\mu\rm m}$, and $M_{\rm molgas}^{\rm dyn}$. Despite the inherent/systematic uncertainties associated with each tracer discussed in Sects.~\ref{sect:introduction} and \ref{sect:COanalysis}, the adoption of standard empirical calibrations and conversion factors to H$_2$ yields excellent agreement among all five molecular gas mass estimates within their respective $1\sigma$ uncertainties, as illustrated in the left panel of Fig.~\ref{fig:Mmolgas-comparison}.

This agreement first suggests that mid-$J$ CO transitions allow to trace the bulk of the molecular gas reservoir of this MS galaxy at $z\sim 4.5$, as predicted by simulations \citep{Vallini+18}. This is likely a consequence of the higher gas densities prevailing in the ISM of high-redshift galaxies compared to their local counterparts \citep{Isobe+23}. Second, the consistency among the various $M_{\rm molgas}$ estimates suggests that, despite its high SFR (Table~\ref{tab:properties}), this solar-metallicity galaxy is characterized by ISM conditions that favour a CO-to-H$_2$ conversion factor close to the Milky Way value rather than the lower $\alpha_{\rm CO}^{\rm SB}$ typically adopted for local starbursts. Indeed, considering $\alpha_{\rm CO}^{\rm SB}$ would result in $M_{\rm molgas}^{\rm CO}$ that is lower by a factor of $\sim 4$ ($\sim 0.64$~dex), leading to a significant discrepancy with the masses inferred from the other tracers. Reconciling such a low $\alpha_{\rm CO}$ with the observed agreement would require all the molecular gas tracers considered here to systematically overestimate $M_{\rm molgas}$ by a similar factor.

The CO excitation ratio measured in DC873756, $r_{5,4}  = L^{\prime}_{\rm CO(5-4)}/L^{\prime}_{\rm CO(4-3)} = 0.53\pm 0.16$, indicates a CO SLED broadly consistent with that of MS galaxies at early cosmic noon \citep[$z=2-2.7$;][]{Boogaard+20}, which we adopt throughout this work (Sect.~\ref{sect:COanalysis}). Comparable values of $r_{5,4}$ have also been reported for three MS galaxies at $z=3.0-3.4$ \citep{Cassata+20}. In contrast, the substantially higher CO excitation typically observed in starbursting DSFGs \citep{Spilker+14} and AGN \citep{Vallini+19}, for which $r_{5,4}$ often exceeds unity, is disfavoured\footnote{The moderate CO excitation provides an additional constraint on the nature of DC873756 and argues against an AGN contribution. This is relevant given the elevated [N\,{\sc ii}]${\lambda 6585}$/H$\alpha$ ratio measured by \citet{Faisst+26}, which, together with the poorly constrained [O\,{\sc iii}]${\lambda 5007}$/H$\beta$ ratio, places the galaxy in the AGN region of the BPT diagram \citep{Baldwin+81}, although other diagnostics favour a star-forming classification (see Fig.~6 of \citealt{Faisst+26}).}. The moderate CO excitation is consistent with the empirical correlation between CO excitation and $\Sigma_{\rm SFR}$ proposed by \citet{Frias+23}, with $\Sigma_{\rm SFR}$ of DC873756 being comparable to that of MS galaxies at cosmic noon, i.e.\ much lower than values typically measured in high-redshift starbursts (see \citealt{Bethermin+23} and \citealp{Accard+25}).

The upper limit on $M_{\rm molgas}^{\rm [CI](1-0)}$ is also consistent with the molecular gas masses derived from the other tracers. The limit was obtained by adopting the excitation conditions and carbon abundance representative of MS galaxies at cosmic noon, analogous to the assumptions underlying the derivation of $M_{\rm molgas}^{\rm CO}$ (Sect.~\ref{sect:CIanalysis}). The non-detection of the [C\,{\sc i}](1--0) line most likely reflects its intrinsic faintness combined with the limited sensitivity of the observations. Indeed, for $M_{\rm molgas}$ inferred from the other tracers, the expected [C\,{\sc i}] emission would only be detected at a modest significance level of $\sim 3-4\sigma$. The upper limit is furthermore consistent with the $L^{\prime}_{\rm [CI](1-0)}$--$L_{\rm IR}$ correlation observed for MS galaxies at cosmic  noon and interpreted as being analogous to the Kennicutt-Schmidt relation \citep{Valentino+18,Valentino+20,Dunne+22}. Nevertheless, the detection of both CO(4--3) and CO(5--4), together with the non-detection of [C\,{\sc i}], may also indicate relatively highly excited ISM conditions, in which the mid-$J$ CO transitions are enhanced with respect to the low-excitation atomic carbon emission, particularly from the $J=1$ level. The CMB is unlikely to be the primary cause of the faint [C\,{\sc i}] emission, since a reduction in line contrast of only $\sim 15$\% is expected for the gas conditions relevant to DC873756 \citep{Zhang+16,Frias+25}, which is insufficient to explain the [C\,{\sc i}] non-detection.

Finally, the close agreement between $M_{\rm molgas}^{\rm CO}$ and $M_{\rm molgas}^{850\mu\rm m}$ in DC873756 indicates that the commonly adopted $\rm DGR \sim 1/100$ \citep{Scoville+16} is appropriate for this solar metallicity galaxy. This is consistent with CO- and dust-based molecular gas masses measured in two other MS galaxies at $z=5.65$ \citep{Pavesi+19} and $z=6.03$ \citep{Zavala+22}, and with simulations predicting only a weak evolution of the DGR in massive, metal-rich star-forming galaxies from $z=0$ to $z\sim5.3$ \citep[e.g.,][]{Li+19,PoppingPeroux22}. In contrast, a significantly lower DGR of $\sim 1/600$ has been reported for the CO-detected REBELS-25 galaxy at $z=7.31$ \citep{Cescon+26}, in line with measurements of other galaxies at $z>6.5$ possibly indicating a turnover in the DGR toward the epoch of reionization \citep{Algera+26}.


\subsubsection{DC818760}
\label{sect:DC818760}

The physical conditions in DC818760 are complex, given that the triple system is undergoing a major merger between the eastern and central galaxies  \citep{Jones+20,Jones+21,Devereaux+24,Lee+25}. We find that $M_{\rm molgas}^{\rm [CII]}$ exceeds $M_{\rm molgas}^{850\mu\rm m}$ by 0.36~dex, a discrepancy larger than the 0.3~dex scatter associated with the fiducial $\alpha_{\rm [CII]}^{\rm Zanella}$ calibration (middle panel of Fig.~\ref{fig:Mmolgas-comparison}). The tentative CO(4--3) detection yields an even lower molecular gas mass estimate (consistent with the upper limit on $M_{\rm molgas}^{\rm CO(5-4)}$), such that $M_{\rm molgas}^{\rm [CII]}$ exceeds $M_{\rm molgas}^{\rm CO(4-3)}$ by 0.64~dex. Figure~\ref{fig:DC818760} yet shows that the CO(4--3) emission originates from the same merging galaxies whose combined [C\,{\sc ii}] luminosity is used to derive $M_{\rm molgas}^{\rm [CII]}$. Thus, the discrepancy is unlikely to arise from a mismatch between the emitting regions traced by the two diagnostics. Adopting an enhanced CO excitation and/or a lower CO-to-H$_2$ conversion factor approaching $\alpha_{\rm CO}^{\rm SB}$ (which could be expected for mergers), would further decrease the inferred $M_{\rm molgas}^{\rm CO(4-3)}$ and increase the discrepancy with $M_{\rm molgas}^{\rm [CII]}$. As a result, the available molecular gas tracers favour an `intrinsic' excess in the [C\,{\sc ii}] luminosity from the merging galaxies, leading to an apparently elevated $M_{\rm molgas}^{\rm [CII]}$ (see Sect.~\ref{sect:Caveats} for further discussion).

An excess in $L_{\rm [CII]}$ has been previously reported in another ALPINE major merger at $z=4.57$ \citep[][see also \citealt{DiCesare+24}]{Ginolfi+20}, as well as in the environments of nearby interacting galaxies \citep{Cormier+12,Appleton+13,VelusamyLanger14}, in agreement with predictions from simulations \citep{Schimek+24}. A systematic excess of $M_{\rm molgas}^{\rm [CII]}$ relative to $M_{\rm molgas}^{\rm CO}$ has also been identified in several highly star-forming quasar host galaxies at $z\gtrsim6$ \citep{Kaasinen+24}. One proposed explanation is the presence of shock-excited [C\,{\sc ii}] emission and/or diffuse ionized gas generated during the merger process, which contribute to $L_{\rm [CII]}$ without tracing the molecular gas reservoir. In this context, \citet{Devereaux+24} reported evidence for diffuse [C\,{\sc ii}] emission surrounding the eastern and central merging galaxies in DC818760, lending further support to the interpretation that the elevated $M_{\rm molgas}^{\rm [CII]}$ results from merger-driven enhancement of the [C\,{\sc ii}] emission.

Finally, the CO excitation ratio $r_{5,4}$ measured in DC873756 naturally explains the absence of even a tentative CO(5--4) detection in DC818760. Indeed, applying it to the $4.3\sigma$ $L^{\prime}_{\rm CO(4-3)}$ measurement of DC818760 yields $L^{\prime}_{\rm CO(5-4)} = 0.31\times10^{10}~\rm K~km~s^{-1}~pc^2$, which is approximately a factor of two below the current $3\sigma$ upper limit (Table~\ref{tab:observations}).

\subsubsection{VC5110377875}
\label{sect:VC5110377875}

Given the morpho-kinematic evidence that VC5110377875 is a rotation-dominated galaxy \citep{Jones+21}, we were able to derive a robust $M_{\rm molgas}^{\rm dyn}$ measurement from the resolved [C\,{\sc ii}] emission \citep{Dessauges+20}. Remarkably, it agrees within $1\sigma$ uncertainty with both $M_{\rm molgas}^{\rm [CII]}$ and $M_{\rm molgas}^{850\mu \rm m}$. It is also consistent with the upper limits on $M_{\rm molgas}^{\rm CO(4-3)}$ and $M_{\rm molgas}^{\rm CO(5-4)}$ as shown in the right panel of Fig.~\ref{fig:Mmolgas-comparison}, which were derived again by assuming a CO SLED and $\alpha_{\rm CO}^{\rm MW}$, representative of MS star-forming galaxies at early cosmic noon ($z=2-2.7$).

The non-detection of the CO(4--3) and CO(5--4) lines may be a consequence of the subsolar metallicity of the galaxy. \citet{Faisst+26} derived a metallicity of $Z\sim 0.5~Z_{\odot}$\footnote{For a solar metallicity $12+\log({\rm O/H})_{\odot} = 8.69$ \citep{Asplund+09}.} in VC5110377875, whereas DC873756 is consistent with a solar metallicity (Table~\ref{tab:properties}). At such a subsolar metallicity,  $\alpha_{\rm CO}$ already increases by more than a factor of two relative to $\alpha_{\rm CO}^{\rm MW}$ in nearby galaxies \citep{Leroy+11,Sandstrom+13,Bolatto+13,Genzel+15,Accurso+17}. Consequently, for a given molecular gas mass, the intrinsic CO(4--3) and CO(5--4) luminosities would be significantly lower than those expected for a solar-metallicity galaxy such as DC873756, potentially explaining their non-detection.

Despite the slightly deeper NOEMA observations of VC5110377875 compared to the other two ALPINE galaxies studied here (Table~\ref{tab:observations}), the [C\,{\sc i}](1--0) line remains undetected. Assuming $X_{\rm [CI]}$ measured for MS galaxies at cosmic noon (Sect.~\ref{sect:CIanalysis}), the resulting upper limit on $M_{\rm molgas}^{\rm [CI]}$ lies approximately 0.3~dex above the molecular gas masses inferred from the other tracers (right panel of Fig.~\ref{fig:Mmolgas-comparison}). This suggests that substantially deeper observations are required to secure a detection of the [C\,{\sc i}] line in VC5110377875. In addition, the subsolar metallicity of the galaxy may increase the weakness of the [C\,{\sc i}] emission, similarly to the CO emission. Indeed, $X_{\rm [CI]}$ has been found to scale approximately linearly with metallicity \citep{Walter+11,HeintzWatson20}. 
\subsection{Caveats}
\label{sect:Caveats}

H$_2$ is the most abundant molecule in the Universe, yet it is rarely used to trace the total molecular gas reservoir because it lacks a permanent dipole moment. Its lowest-energy quadrupole transitions probe only the warm molecular phase ($\gtrsim 510$~K), which represents a small fraction ($1\%-30$\%) of the total H$_2$ reservoir \citep{Roussel+07}, while the majority of the star-forming molecular gas resides in the cold ($\lesssim 100$~K) phase. Thus, alternative tracers, including CO, [C\,{\sc ii}], [C\,{\sc i}] lines, dust continuum, and, more recently, the dynamical mass, are commonly employed to infer the total cold molecular gas content. However, each of these tracers is subject to its own systematic uncertainties and relies on empirical calibrations and conversion factors to recover the underlying H$_2$ mass (see Sects.~\ref{sect:introduction}, \ref{sect:COanalysis}, and \ref{sect:CIanalysis}). Our small sample of three [C\,{\sc ii}]-luminous ALPINE galaxies provides one of the first cross-validation studies of six tracers at $z\sim 4.5$. 

The CO rotational emission lines are direct and among the most commonly used tracers of the cold molecular gas in nearby galaxies. Therefore, we specifically targeted the CO lines in the three MS galaxies at $z\sim 4.5$ to provide a reference for their molecular gas estimates derived from different tracers. At their high redshifts, however, only mid-$J$ CO transitions are expected to remain easily accessible given their mildly reduced contrast against the elevated CMB temperature. While mid-$J$ CO lines transitions do not necessarily trace the bulk of the cold molecular gas reservoir in nearby galaxies, simulations predict that they should do in high-redshift systems \citep{Vallini+18} because of the higher gas densities prevailing in their ISM \citep{Isobe+23}.

Our search for CO(4--3) and CO(5--4) emission in the three galaxies provides the following insights, while acknowledging that the small sample limits the extent to which these results can be generalized. 
First, our results are consistent with the well-established dependence of CO emission on ISM metallicity \citep[e.g.][]{Leroy+11,Sandstrom+13,Bolatto+13}, with CO emission becoming increasingly difficult to detect in subsolar media. This likely explains the CO non-detection of VC5110377875 with its half-solar metallicity (Sect.~\ref{sect:VC5110377875}), and may also contribute to the CO non-detection of the western galaxy in the DC818760 system (Fig.~\ref{fig:DC818760}).
Second, converting CO luminosities into H$_2$ masses requires assumptions regarding both the CO SLED and $\alpha_{\rm CO}$, which depend on ISM properties including not only metallicity but also density and temperature \citep{Bolatto+13,CarilliWalter13}. The CO excitation and the conversion factor commonly adopted for MS galaxies at cosmic noon appear to remain applicable at $z\sim 4.5$, given that they yield molecular gas estimates for DC873756 that agree with those inferred from other independent tracers, while also remaining consistent for the CO upper limits obtained for VC5110377875 (Fig.~\ref{fig:Mmolgas-comparison}). Moderately excited molecular gas conditions and a CO-to-H$_2$ conversion factor close to $\alpha_{\rm CO}^{MW}$ are further supported by the studies of \citet{Pavesi+19} and \citet{Cescon+26}, who have reached similar conclusions for two MS galaxies at $z=5.65$ and $z=7.31$, respectively.
Third, although mid-$J$ CO transitions generally probe denser and more excited gas in nearby galaxies, the agreement among the molecular gas tracers in DC873756 suggests that they may trace the bulk of the molecular gas reservoir in at least some MS galaxies at high redshift.

Taken together, these results suggest that mid-$J$ CO lines are promising tracers of molecular gas in near-solar metallicity MS galaxies at $z\sim4.5$ and above, and may provide reliable molecular gas masses when adopting the CO SLED and $\alpha_{\rm CO}^{\rm MW}$ commonly assumed for MS galaxies at cosmic noon, although larger samples are required to assess the general applicability of these conclusions. In contrast, CO emission may become difficult to detect in sub-solar media, already at metallicities comparable to that of VC5110377875 ($Z\sim 0.5~Z_\odot$). This represents an important caveat of CO as a molecular gas tracer, since JWST studies indicate that most of the galaxies at $z = 4-10$ have $Z\lesssim 0.5~Z_\odot$ across a broad stellar mass range \citep{Nakajima+23,Sarkar+25}.

[C\,{\sc ii}] is by far the brightest and therefore the most readily accessible FIR emission line at $z>4$, as demonstrated by the ALPINE \citep{LeFevre+20,Bethermin+20} and REBELS \citep{Bouwens+22} surveys, and remains detectable down to metallicities of $Z\sim 0.2~Z_\odot$ \citep{Faisst+26}. This makes [C\,{\sc ii}] an attractive molecular gas tracer in the high-redshift Universe despite its complex multiphase origin, as the line can be excited by collisions with electrons, neutral hydrogen atoms, and molecular hydrogen. Nevertheless, because the brightest [C\,{\sc ii}] emission generally arises from dense PDRs associated with the outer layers of giant molecular clouds \citep{HollenbachTielens99,Wolfire+22}, an analytical relation between the [C\,{\sc ii}] surface brightness of PDRs and the cold H$_2$ column density exists \citep{Ferrara+19}. This provides a physical basis for the empirical $L_{\rm [CII]}$--$M_{\rm molgas}$ calibration observed in both nearby and high-redshift galaxies.

Several observational \citep[e.g.,][]{Accurso+17,Zanella+18,Madden+20,Rizzo+21, Ramambason+24,Kaasinen+24, Zhao+24} and simulation \citep[e.g.,][]{Vizgan+22,Casavecchia+25, Vallini+25} studies have tried to calibrate the $L_{\rm [CII]}$--$M_{\rm molgas}$ relation. Most measurements are consistent with the fiducial $\alpha_{\rm [CII]}^{\rm Zanella} = 31\pm 0.2~M_{\odot}~L_{\odot}^{-1}$ within a scatter of $\sim 0.6$~dex, although significantly lower values ($\alpha_{\rm [CII]}\approx 4.5-7~M_{\odot}~L_{\odot}^{-1}$) have been reported for gravitationally lensed dusty star-forming galaxies at $z\sim 4.5$ \citep{Rizzo+21} and quasar host galaxies at $z\sim 6$  \citep{Kaasinen+24}. Observational studies have not identified a clear dependence of $\alpha_{\rm [CII]}$ on redshift, stellar mass, distance from the MS, or metallicity, whereas recent simulations predict dependencies on metallicity \citep{Vallini+25} and redshift \citep{Casavecchia+25}. Analogous to routine practices for CO, adopting a constant $\alpha_{\rm [CII]}$ conversion factor is therefore expected to provide a first-order estimate of the molecular gas mass. The agreement between the molecular gas masses inferred for DC873756 from six different tracers, including the CO measurements, lends further support to the use of [C\,{\sc ii}] emission with the fiducial $\alpha_{\rm [CII]}^{\rm Zanella}$ for estimating $M_{\rm molgas}$.

The impact of local physical conditions on scales below $\sim 10$~pc has recently been investigated using zoom-in simulations \citep[e.g.,][]{Gurman+24,Vallini+25,Accard+26}. These studies in fact show that adopting a spatially-resolved $\alpha_{\rm [CII]}$ linked to the local gas density and metallicity variations improves the accuracy of [C\,{\sc ii}]-based molecular gas estimates. The merging system DC818760 may illustrate the limitations of the global $\alpha_{\rm [CII]}$ approximation. While the observed excess of $M_{\rm molgas}^{\rm [CII]}$ could reflect additional [C\,{\sc ii}] emission powered by merger-driven shocks and/or diffuse ionized gas as discussed in Sect.~\ref{sect:DC818760}, it could also arise, at least in part, from adopting the fiducial $\alpha_{\rm [CII]}^{\rm Zanella}$. Major mergers are expected to exhibit particularly strong spatial variations in the ISM density, and the simulations of \citet[][see their Fig.~4]{Vallini+25} predict a steep anti-correlation of the local $\alpha_{\rm [CII]}$ with gas density. In this scenario, applying a global $\alpha_{\rm [CII]}^{\rm Zanella}$ may overestimate the molecular gas mass associated with the densest regions, thereby contributing to the elevated $M_{\rm molgas}^{\rm [CII]}$ inferred for DC818760.

\section{Conclusions}
\label{sect:conclusions}

In this work, we searched for CO(4--3), CO(5--4), and [C\,{\sc i}](1--0) emission lines in three of the most [C\,{\sc ii}]-luminous ALPINE galaxies at $z\sim 4.5$ to assess the detectability of these lines in high-redshift MS galaxies and evaluate the reliability of [C\,{\sc ii}] emission as a molecular gas tracer now widely used in high-redshift studies. These galaxies were selected because their high [C\,{\sc ii}] luminosities implied large molecular gas reservoirs based on the $L_{\rm [CII]}$--$M_{\rm molgas}$ relation of \citet{Zanella+18} and other molecular gas tracers (dust continuum and [C\,{\sc ii}]-based dynamical masses; \citealt{Dessauges+20}). The three selected galaxies probe the massive and high SFR end of the ALPINE galaxy distribution in the MS plane (Fig.~\ref{fig:MS}). We have also, on purpose, considered three galaxies with different ISM morpho-kinematic properties as derived from resolved [C\,{\sc ii}] emission (dispersion-dominated, major merger, and rotation-dominated; \citealt{LeFevre+20} and \citealt{Jones+21}) to investigate diverse ISM conditions of MS galaxies at $z\sim 4.5$. This work presents one of the rare studies where up to six different molecular gas tracers are cross-compared in three galaxies representative of the MS galaxy population at the end of the cosmic reionisation era. 

CO(4--3) and CO(5--4) are both detected in the solar metallicity galaxy DC873756 at high significance levels ($10.5\sigma$ and $8.6\sigma$, respectively), while only a tentative CO(4--3) detection ($4.3\sigma$) is obtained for the merging system DC818760, and no CO emission is detected in the half-solar metallicity galaxy VC5110377875 (Fig.~\ref{fig:spectra}). The [C\,{\sc i}](1--0) line remains undetected in all three galaxies likely because of its intrinsic faintness combined with the limited sensitivity of the NOEMA observations.

In DC873756, the molecular gas masses inferred from the CO lines, dust continuum, [C\,{\sc ii}] emission, and [C\,{\sc ii}]-based dynamical mass agree within their $1\sigma$ uncertainties despite the different systematics inherent to each tracer (Table~\ref{tab:properties} and Fig.~\ref{fig:Mmolgas-comparison}). This consistency suggests that, at least for some near-solar metallicity MS galaxies at $z\sim4.5$, the CO SLED and $\alpha_{\rm CO}^{\rm MW}$ commonly adopted for MS galaxies at cosmic noon \citep[$z= 2-2.7$;][]{Boogaard+20} may remain applicable, and that mid-$J$ CO transitions may trace a substantial fraction of the molecular gas reservoir, as theoretically expected because of the higher ISM densities of galaxies toward the epoch of reionisation \citep{Vallini+18}. Conversely, the non-detection of CO in the $Z\sim 0.5~Z_{\odot}$ galaxy VC5110377875 is consistent with CO emission becoming increasingly difficult to detect at lower metallicities. These conclusions are further supported by the studies of \citet{Pavesi+19} and \citet{Cescon+26}, which likewise favour moderately excited molecular gas and a CO-to-H$_2$ conversion factor close to $\alpha_{\rm CO}^{\rm MW}$ in two CO-detected MS galaxies at $z=5.65$ and $z=7.31$, the latter having a near-solar metallicity \citep[$Z\sim 0.85~Z_{\odot}$;][]{Rowland+26} This is in line with the emerging picture that CO seems to remain detectable in metal-rich MS galaxies at high redshift.

The agreement among the different molecular gas tracers in DC873756 and VC5110377875 also lends support to the empirical $L_{\rm [CII]}$--$M_{\rm molgas}$ relation and the use of [C\,{\sc ii}] emission as a first-order estimator of molecular gas masses in high-redshift MS galaxies. At the same time, the apparent [C\,{\sc ii}] excess in the merging system DC818760 illustrates that [C\,{\sc ii}] may not provide an unbiased molecular gas estimate in disturbed environments, either because of additional emission associated with shocks and diffuse ionized gas, as reported in nearby interacting galaxies \citep{Cormier+12,Appleton+13,VelusamyLanger14} and in another ALPINE major merger at $z=4.57$ \citep{Ginolfi+20,DiCesare+24}, or because a global $\alpha_{\rm [CII]}$ cannot account for the strong spatial variations in ISM conditions expected in mergers. Recent zoom-in simulations show that accounting for local ($\sim 10$~pc) variations in gas density and metallicity through a spatially-resolved $\alpha_{\rm [CII]}$ improves the accuracy of [C\,{\sc ii}]-based molecular gas estimates \citep[e.g.,][]{Gurman+24,Vallini+25,Accard+26}. However, achieving such a small spatial resolution remains beyond the capabilities of current observational facilities for large galaxy samples, making the use of a global $\alpha_{\rm [CII}]$ for now unavoidable.

While necessarily limited by the small sample size, this study provides one of the rare observational cross-checks of up to six molecular gas tracers in MS galaxies at $z\sim 4.5$. It brings a first assessment of the reliability of the now widely used [C\,{\sc ii}] emission as a molecular gas tracer in $z>4$ galaxies against CO lines that have long served as the benchmark for molecular gas studies in the nearby Universe. Confirming this finding will require larger samples with both CO- and [C\,{\sc ii}]-based molecular gas estimates at high redshift. Beyond this, our results also motivate renewed CO studies of high-redshift MS galaxies. The strong CO(4--3) and CO(5--4) detections in the near-solar metallicity galaxy DC873756 suggest that mid-$J$ CO emission can be readily detected in such systems and provide molecular gas masses consistent with independent tracers. This prospect is particularly timely given the advent of ALMA Band~1, which enables observations of the CO(4--3) transition up to $z\sim 12$, together with the rapidly growing number of JWST metallicity measurements for galaxies at $z>4$ \citep[e.g.,][]{Schaerer+22,Nakajima+23,Sanders+24,Sanders+26}. These developments make it possible to preselect near-solar metallicity galaxies for CO follow-up observations, paving the way for systematic studies of molecular gas in MS star-forming galaxies during the first $\sim 1.5$~Gyr of cosmic history. 



\begin{acknowledgements}
This work is based on observations carried out under project number W22DW with the IRAM NOEMA Interferometer. IRAM is supported by INSU/CNRS (France), MPG (Germany), and IGN (Spain).
This paper makes use of the ALMA data: ADS/JAO.ALMA\#2017.1.00428.L and 2019.1.00226.S. ALMA is a partnership of ESO (representing its member states), NSF (USA) and NINS (Japan), together with NRC (Canada), NSC and ASIAA (Taiwan), and KASI (Republic of Korea), in cooperation with the Republic of Chile. The Joint ALMA Observatory is operated by ESO, AUI/NRAO and NAOJ. 
YF acknowledge supports by JSPS KAKENHI Grant Numbers JP22K21349 and JP23K13149.
GCJ acknowledges support by the Science and Technology Facilities Council (STFC), by the ERC through Advanced Grant 695671 ``QUENCH'', and by the UKRI Frontier Research grant RISEandFALL.
CA acknowledges support from the Interdisciplinary Thematic Institute IRMIA++ (ITI 2021–2028, University of Strasbourg, CNRS, Inserm) funded by IdEx Unistra (ANR-10-IDEX-0002) and SFRI-STRAT’US (ANR-20-SFRI-0012) under the French Investments for the Future program.
\end{acknowledgements}

\begin{appendix}
\section{DC873756 and DC818760 images}

\begin{figure}[h]
\hspace{0.3cm}
\includegraphics[width=0.48\textwidth,clip]{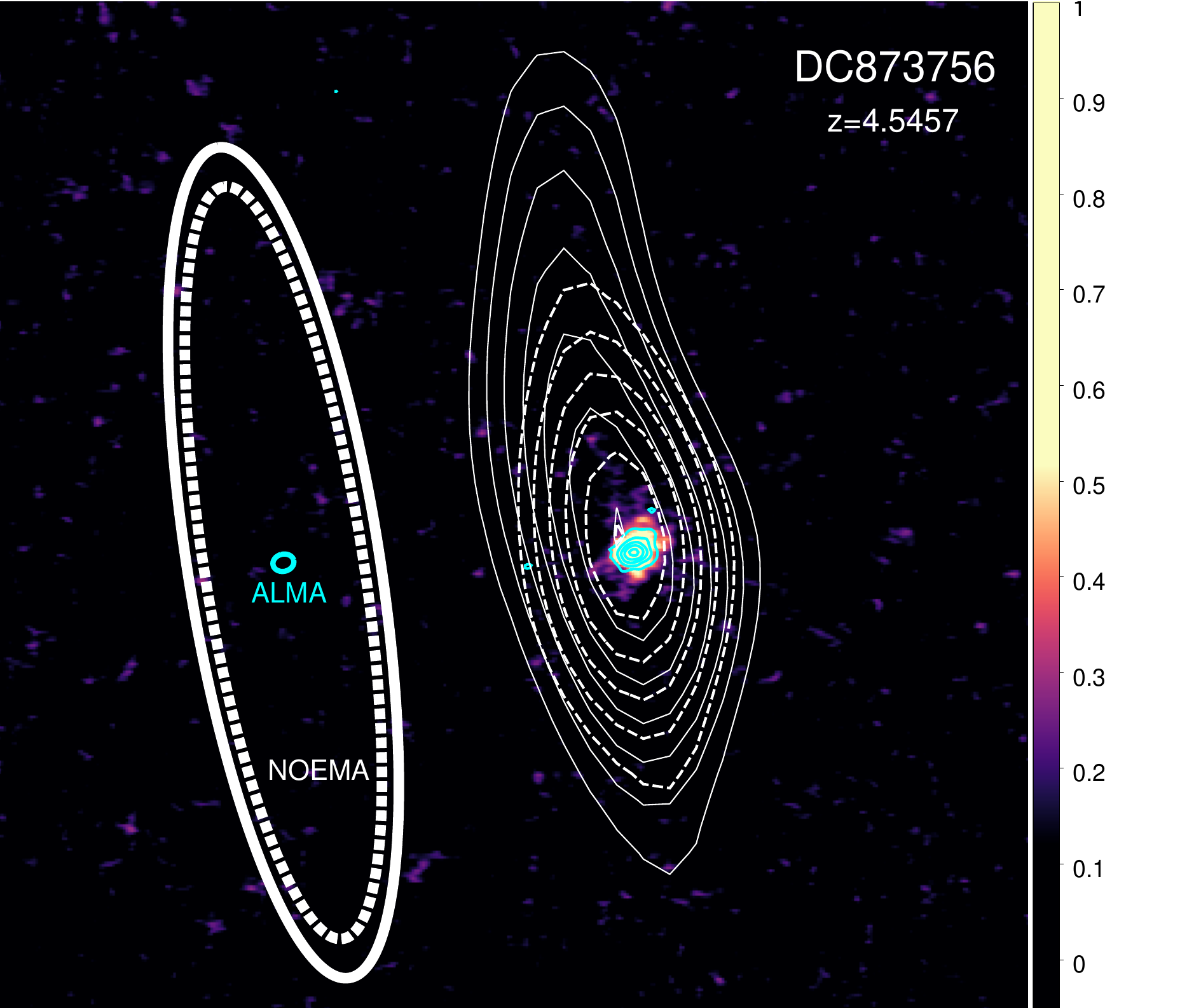}
\caption{The background shows the velocity-integrated [C\,{\sc ii}] moment-0 map, with the color scale in units of Jy~beam$^{-1}$ km~s$^{-1}$, of the dispersion-dominated DC873756 galaxy, obtained by \citet[][see also \citealt{Devereaux+24}]{Bethermin+23} from the concatenation of the ALMA high ($0.15\arcsec$), medium ($0.3\arcsec$), and low ($0.9\arcsec$) resolution observations. The corresponding ALMA synthesised beam size of $0.32\arcsec \times 0.26\arcsec$ is shown by the cyan ellipse on the left. The ALMA rest-frame $\sim 158~\mu$m dust continuum contours are overlaid in cyan starting from $4\sigma$ significance level in steps of $4\sigma$. 
The NOEMA CO(4--3) and CO(5--4) emissions are shown with white solid and dashed contours, respectively, starting from $4\sigma$ significance level in steps of $1\sigma$. The corresponding NOEMA synthesised beams are the white solid and dashed ellipses with sizes of $12.7\arcsec \times 2.9\arcsec$ and $11.5\arcsec \times 2.4\arcsec$. East is left and north is up.}
\label{fig:DC873756}
\end{figure}

\begin{figure}[h]
\hspace{0.3cm}
\includegraphics[width=0.48\textwidth,clip]{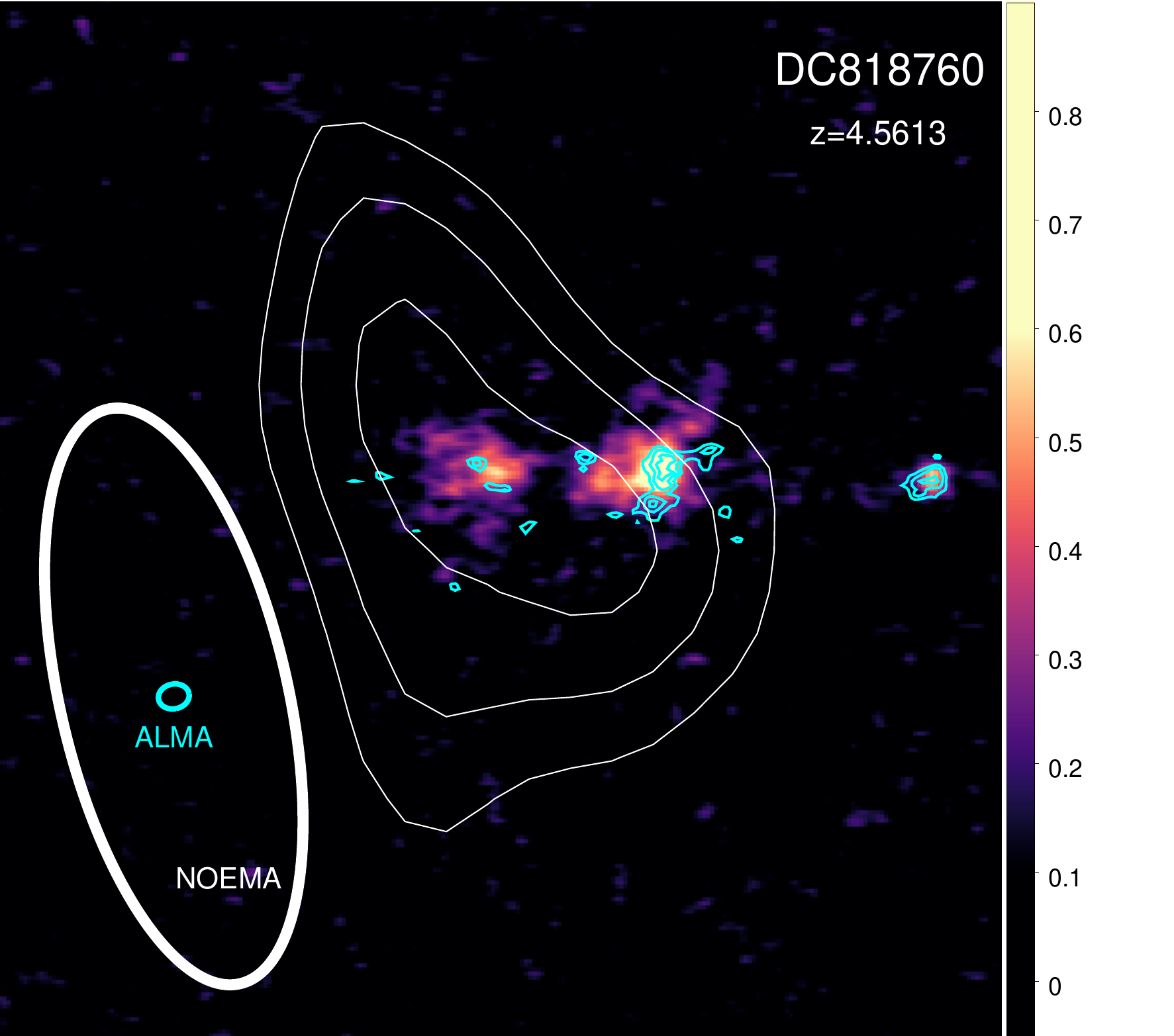}
\caption{Same as in Fig.~\ref{fig:DC873756} for the triple galaxy system DC818760.
The corresponding ALMA synthesised beam size of $0.30\arcsec \times 0.23\arcsec$ is shown by the cyan ellipse on the left. The eastern and central galaxies are experiencing a major merger \citep{Jones+20,Jones+21,Romano+21,Devereaux+24,HerreraCamus+25}. The ALMA rest-frame $\sim 158~\mu$m dust continuum contours are overlaid in cyan starting from $3\sigma$ significance level in steps of $1\sigma$. 
The NOEMA CO(4--3) emission is shown with white contours at 2, 3, and $4\sigma$ levels and comes from the eastern and central galaxies. The NOEMA synthesised beam is the white ellipse with a size of $5.7\arcsec \times 2.2\arcsec$. East is left and north is up.}
\label{fig:DC818760}
\end{figure}

\end{appendix}

\end{document}